\documentclass[%
 prl,
superscriptaddress,
twocolumn,
 amsmath,amssymb,
 aps,
]{revtex4-1}

\usepackage{amssymb}
\usepackage{amsmath}
\usepackage{braket}
\usepackage{verbatim}
\usepackage{xcolor}
\usepackage{wasysym}
\usepackage{enumitem}

\usepackage{graphicx}
\usepackage{dcolumn}
\usepackage{bm}
\usepackage{hyperref}
\usepackage[normalem]{ulem}

\renewcommand{\vec}{\mathbf}

\begin{document}

\title{Light-harvesting with guide-slide superabsorbing condensed-matter nanostructures}

\author{W. M.  Brown}
\email{Electronic address: wmb1@hw.ac.uk}
\author{E. M. Gauger}%
\email{Electronic address: e.gauger@hw.ac.uk}
\affiliation{%
 SUPA, Institute of Photonics and Quantum Sciences,\\
 Heriot-Watt University, Edinburgh, EH14 4AS, United Kingdom
}%

\date{\today}

\begin{abstract}
We establish design principles for light-harvesting antennae whose energy capture scales superlinearly with system size. Controlling the absorber dipole orientations produces sets of `guide-slide' states which promote steady-state superabsorbing characteristics in noisy condensed-matter nanostructures. Inspired by natural photosynthetic complexes, we discuss the example of ring-like dipole arrangements and show that, in our setup, vibrational relaxation enhances rather than impedes performance. Remarkably, the superabsorption effect proves robust to $\mathcal{O}(5 \%)$ disorder simultaneously for all relevant system parameters, showing promise for experimental exploration across a broad range of platforms.
\end{abstract}

\maketitle

\textit{Introduction} -- 
Photosynthesis powers most life on Earth~\cite{Blankenship} and may provide templates for artificial light-harvesting~\cite{scholes11,Romero:2017aa,Scholes:2017aa}. 
Recent years have seen several proposals for enhancing the performance of quantum photocells beyond the venerable Shockley-Queisser limit of traditional photovoltaic devices~\cite{Shockley1961}. Many of these aim to prevent recombination once a photon has been absorbed, e.g.~through interference in multi-level systems~\cite{Scully:2010aa, Scully2011} or as dark-state protection with multiple interacting dipoles~\cite{Creatore2013a, Yamada2015a, Fruchtman2016, Zhang2015}. Optical ratcheting \cite{Higgins2017, Zhang:2016aa} may offer further advantages by allowing an antenna to keep absorbing whilst being immune to recombination. Recent work underlines the importance of expanding beyond the single excitation subspace even in the presence of exciton-exciton annihilation~\cite{Hu_2018,Higgins2017}. Further, it has also been proposed that coherent vibrations~\cite{Killoran2015, Stones2017}, as well as excitonic coherences~\cite{Tomasi2018} could be beneficial.

An alternative approach to improving the performance of light harvesters would be to enhance the effective optical absorption rate. In 1954 Dicke predicted the phenomenon of superradiance, where $N$ atoms exhibit a collectively-enhanced `greater-than-the-sum-of-its-parts' emission rate $\propto N^2$~\cite{Dicke1954a}. The possibility of harnessing the time-reversed phenomena has recently been proposed: (slightly) lifting degeneracies through symmetric dipolar interactions allows environmental control which temporarily keeps the system in a regime where it displays `superabsorption'~\cite{Higgins2014a}. This does not occur in natural systems but was inspired by photosynthetic ring antenna surrounding a reaction centre that converts optically-created excitations into useful chemical energy~\cite{Law2004,Sumi2001}. In natural rings, dipoles tend to align tangentially around the ring~\cite{Hu1997}, limiting the overall light-matter coupling but also allowing for inclusion of mechanisms for safeguarding against photo-damage~\cite{Takahashi2011}. On the other hand, for artificial systems configurations with a stronger collective dipole may prove more advantageous, as we explore in the following. 

A very recent experimental study reports the observation of superabsorption with atomic systems~\cite{Yang2019}, providing a strong motivation for tackling the challenge of harnessing collectively enhanced absorption in condensed matter and particularly molecular structures.
Encouragingly, molecular rings displaying coherence effects can be synthesised~\cite{OSullivan2011a, Butkus:2017aa}, and cooperative dipole behaviour remains attainable under strong dephasing~\cite{Venkatesh2018}. Delocalised excitonic states of the kind necessary for the proposals listed above may also occur in stacks of rings which self assemble into symmetric nanotubes~\cite{Gulli2019,Lohner2019}.

\begin{figure}
\centering
	\includegraphics[width=0.47\textwidth]{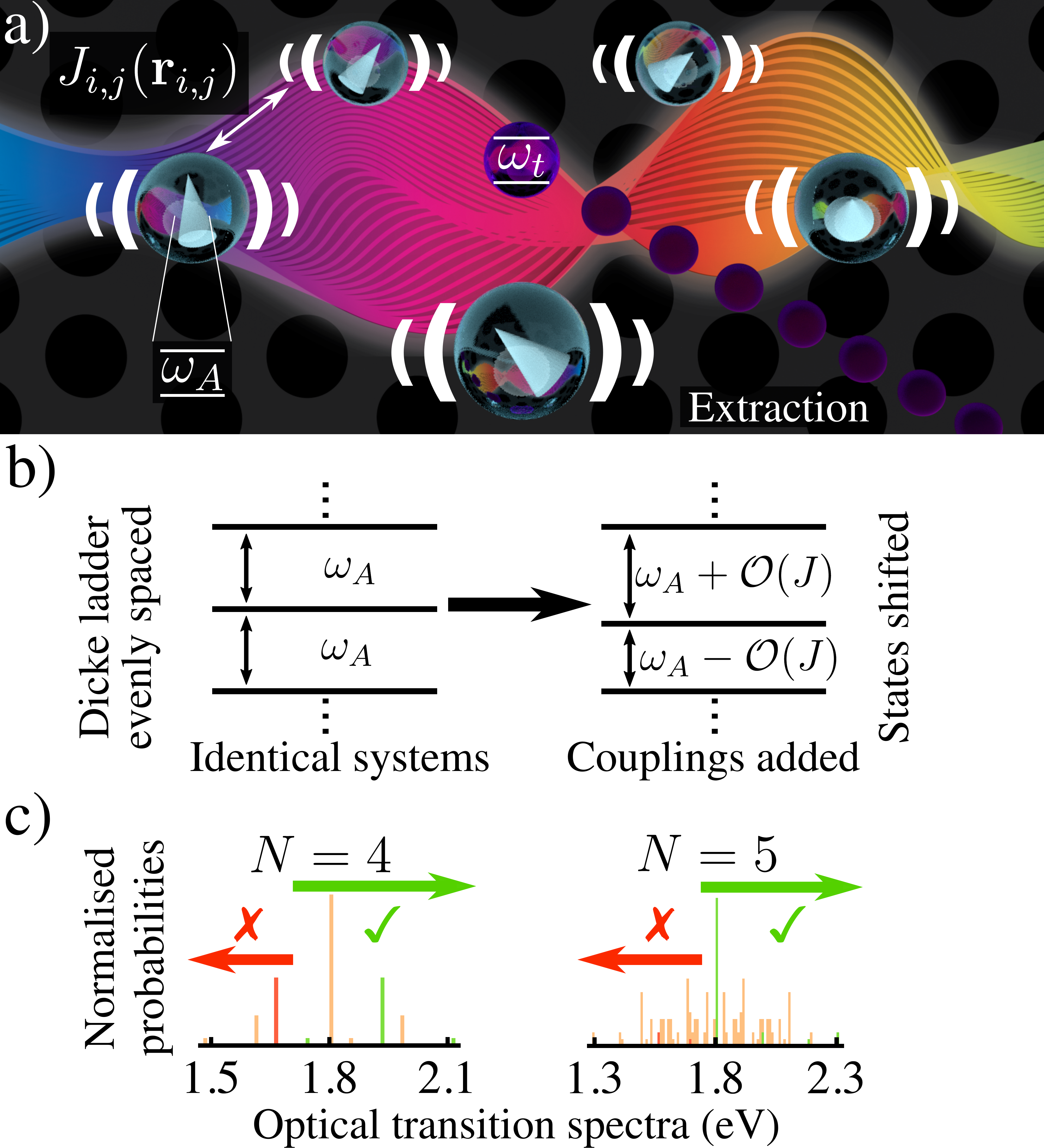}
\caption{a) Artistic depiction of a guide-slide superabsorber: A ring of skewed optical dipoles interacting with a shared photon environment, local phonon baths, and a central extracting trap. A photonic crystal suppresses interaction with certain optical modes. b) Couplings $J_{i,j}$ between dipoles cause a perturbation to the Dicke ladder, primarily lifting the degeneracy of ladder rung spacings. c) Histogram of transition frequencies for coupled dipoles: there is no overlap between undesirable modes to be suppressed (red) and light-harvesting target modes (green) irrespective of ring size~\cite{SI}.}
\label{fig:glossy}
\end{figure}

In this Letter, we examine the potential of condensed-matter nanostructures to operate as collectively-enhanced quantum photocells. Finding that the geometry of Ref.~\cite{Higgins2014a} fails in the presence of vibrational relaxation, we focus on a different means of achieving a cooperative advantage. We define a {\it guide-slide superabsorber} to be a collection of optical dipoles possessing the following properties:

\begin{enumerate}[label=\Roman*.]
\item A ladder of excitation manifolds, each with rapid relaxation to a well-defined lowest energy state.
\item Collectively-enhanced optical rates linking the lowest energy states of adjacent manifolds.
\item Spectral selectivity allowing suppression of optical decay below an enhanced target transition.
\end{enumerate}

\textit{Model} -- We consider a ring of $N$ optical dipoles (see Fig.~\ref{fig:glossy}a), each modelled as a degenerate two-level system (2LS) with transition energy $\omega_A = 1.8$~eV ($\hbar=1$), near the peak of the solar power spectrum. Letting $N$ uncoupled dipoles interact collectively with the electromagnetic environment leads to a Dicke ladder with $N+1$ equally spaced `rungs' separated by steps of $\omega_A$.
Each rung represents a collective state for a different number of shared excitations, and optical rates near the middle of this ladder are collectively-enhanced and proportional to $N^2$~\cite{Dicke1954a, Gross1982}.

Interactions between dipoles perturb this picture, however, for moderate coupling strength, the system retains a ladder of eigenstates connected by enhanced optical transitions (Fig.~\ref{fig:glossy}b). The rungs are no longer evenly spaced, instead we obtain a chirped profile with a frequency increment determined by the strength of the dipolar couplings~\cite{Higgins2014a}. This is accompanied by a partial re-distribution of the oscillator strength away from ladder transitions, and a richer optical spectrum (Fig.~\ref{fig:glossy}c).

For closely spaced absorbers, dipolar interactions arise naturally from the `cross Lamb-shift' (resonant F\"orster) terms in the many-body quantum optical master equation~\cite{Varada1992,Curutchet2017, Breuer2002},
\begin{equation}
J_{i,j}(\vec{r}_{i,j})=\frac{1}{4\pi\epsilon_0 |\vec{r}_{i,j}|^3}\bigg(\vec{d}_i\cdot\vec{d}_j-\frac{3(\vec{r}_{i,j}\cdot\vec{d}_i ) ( \vec{r}_{i,j}\cdot\vec{d}_j)}{|\vec{r}_{i,j}|^2}\bigg) ~, 
\nonumber
\end{equation}
where $\vec{r}_{i,j}$ is the vector linking the two dipoles $i,j$, and $\vec{d}_i$ is the dipole moment at site $i$ whose strength is related to the natural lifetime $\tau_L$ of an isolated 2LS by $|\vec{d}| = \sqrt{3\pi \epsilon_0 \tau_L^{-1} c^3 / \omega_A^3}$ \cite{Agarwal1974,Shatokhin2018}. 

Employing the usual definition of site-defined Pauli operators, the ring Hamiltonian reads 
\begin{equation}
\label{eq:Hsysring}
\hat{H}_{\text{ring}}=\omega_A\sum_{i=1}^N \hat{\sigma}^z_i+\sum_{i,j=1}^N J_{i,j}(\vec{r}_{i,j})(\hat{\sigma}^+_i\hat{\sigma}^-_j+\hat{\sigma}^-_i\hat{\sigma}^+_j) ~.
\end{equation}
Assuming a ring diameter that is small compared to relevant photon wavelengths ($ \sim 2 \pi c / \omega_A$) we may neglect phase factors in the coupling elements and write the optical interaction Hamiltonian as
\begin{equation}
\label{eq:Hoint}
\hat{H}_{I,\text{opt}}=\sum_{i=1}^N \vec{d}_i\hat{\sigma}_i^x \otimes \sum_k f_k (\hat{a}_k+\hat{a}_k^\dagger) ~,
\end{equation}
where $f_k$ and $\hat{a}^{(\dagger)}_k$ are, respectively, the coupling strength and annihilation (creation) operator for the optical mode $k$~\cite{Breuer2002,Higgins2014a}.

We begin by contrasting a ring of dipoles that are all perpendicular to the plane of the ring ($||$-SA) to the guide-slide setup (GS-SA), where each dipole has been tilted `sideways' by $ \theta_{\text{zen}} = \pi/4$ (see Fig.~\ref{fig:glossy}a and Supplementary Information (SI)~\cite{SI}). The purpose of this tilting is to flip the sign of nearest-neighbour $J_{i,j}$ terms in Eq.~\eqref{eq:Hsysring} whilst preserving a substantial collective dipole strength $\vec{D} = \sum_i^N \vec{d}_i$.

\begin{figure*}
\centering
	\includegraphics[width=0.75\textwidth]{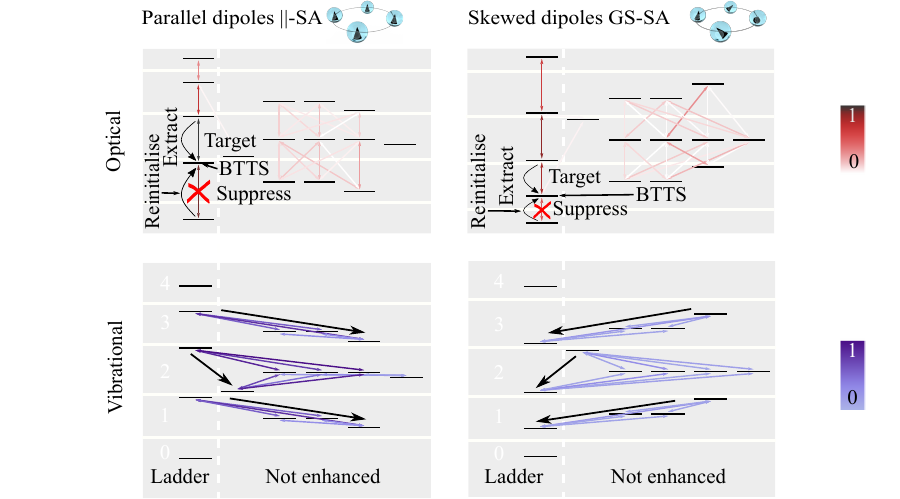}
\caption{Process diagram showing optical (red, top) and vibrational (blue, bottom) processes linking the eigenstates for a quadmer in the $||$-SA (left) and GS-SA (right) setup; normalised colour denotes relative strengths. The states are organised in manifolds corresponding to the number of excitations in the system; the manifolds are visually separated by gaps in the background shading (for clarity energetic intra and inter manifold spacings are not to scale). Optical processes link different manifolds whereas vibrational ones act `sideways'. Superabsorption is achieved by pairing a reinitialisation process with a suppression of optical decay below the collectively-enhanced `target' transition. Black arrows show how the GS-SA system is stabilised by vibrational relaxation, whereas for $||$-SA phonons are detrimental, pulling the system away from superabsorbing states.
}
\label{fig:panels}
\end{figure*}

Transforming Eq.~\eqref{eq:Hoint} into the eigenbasis of Eq.~\eqref{eq:Hsysring} reveals connections between ring eigenstates by optical processes. This is shown in the top panels of Fig.~\ref{fig:panels} for the case of an $N=4$ ring (quadmer) for both the $||$-SA and the GS-SA configuration, in the latter case only for the `collective dipole' $\vec{D}$ (but a full map is given in the SI~\cite{SI}).
States linked by collectively-enhanced transitions -- which overlap significantly with Dicke ladder states ({\it cf.}~Fig.~\ref{fig:glossy}b) -- are shown in the left-most columns, and we henceforth refer to these as {\it ladder states}. The colouring indicates the relative strengths of the transition matrix elements~\footnote{These are symmetric with respect to absorption and emission (detailed balance follows from asymmetric rates), and they include the cubic frequency dependence of the free space spectral density of optical modes~\cite{Breuer2002}.}. It is important to note that $||$-SA ladder states possess the highest energy in their respective excitation manifolds, whereas the opposite is the case for GS-SA.

Reflecting the condensed-matter nature of typical nanostructures, we introduce local vibrational baths which are generically coupled to each 2LS via~\cite{mahan00}:
\begin{equation}
\label{eq:Hpint}
\hat{H}_{I,\text{vib}}=\sum_{i=1}^N \hat{\sigma}_i^z \otimes \sum_q g_{i,q} (\hat{b}_{i,q}+\hat{b}_{i,q}^\dagger) ~,
\end{equation}
where $g_{i,q}$ and $\hat{b}^{(\dagger)}_{i,q}$ are, respectively, the coupling strength and annihilation (creation) operator for the phonon mode $q$ for the bath associated with site $i$~\cite{Breuer2002,Higgins2017}.

The bottom panels of Fig.~\ref{fig:panels} show the resulting phonon processes linking quadmer eigenstates for $||$-SA and GS-SA, colour-coded to indicate relative strength \footnote{This is based on an structureless Ohmic phonon spectral density.}. Phonon processes only link states within the same excitation manifold and typically occur on a much faster timescale than optical processes. Obeying detailed balance, vibrational relaxation preferentially occurs `downwards' in energy~\cite{Gauger2010, Ramsay2011}. For $||$-SA this implies that phonons exert a pull away from the ladder linked by enhanced optical processes, ruining the suitability of this configuration for steady-state superabsorption (full details in the SI~\cite{SI}). By contrast, for GS-SA vibrational dissipation guides the system back to this ladder. A ring of skewed dipoles therefore meets all requirements I-III, and from hereon we shall focus on this configuration.

Wishing to exploit the large oscillator strength at the ladder mid-point, we define a `target transition' with frequency $\omega_{\text{good}}$. As states near the middle of the ladder are not naturally substantially populated, additional measures are required for pinning the system at the `bottom-of-the-target-transition-state' (BTTS). Let the frequency of the ladder transition immediately below the BTTS be $\omega_{\text{bad}}$. Following Ref.~\cite{Higgins2014a}, spectral selectivity permits the application of environmental suppression of certain optical modes. Given a sufficient gap between $\omega_{\text{good}}$ and $\omega_{\text{bad}}$ (which is proportional to $\vert J_{i,j} \vert$), optical decay from the BTTS can be suppressed using a photonic band-gap environment. Fig.~\ref{fig:glossy}c shows that the transition frequencies of upwards absorption (green) and downwards emission (red) processes from the BTTS are indeed well-separated, meaning suppression of a single band of wavelengths will be effective: for GS-SA one may thus suppress all modes with frequency $\omega< \left( \omega_{\text{good}}+\omega_{\text{bad}} \right)/2$ (see Fig.~\ref{fig:glossy}c). A detailed discussion of the suppression of optical processes and separation of desirable and undesirable frequencies, including for $||$-SA, 
is given in the SI~\cite{SI}.

Suppressing emission from the BTTS prolongs the system's proclivity for remaining in this state, keeping it primed for superabsorbing behaviour. However, an initially excited ring may nonetheless relax towards the ground state at longer times. Addressing this issue requires a physical excitation process which keeps pushing the ring system towards the centre of the ladder. This can be accomplished in two distinct ways: first, the spectrally selective suppression naturally allows the system to passively ratchet itself up to the BTTS, and we found this works well for higher levels of optical suppression. Second, for a lower degree of suppression, an actively controlled excitation process can be added.
The latter can be modelled, in an idealised fashion, as incoherent pumping into the BTTS with rate $\gamma_r$ from all ladder states below the BTTS (or alternatively and more realistically using potentially loss-inducing mechanisms like pumping higher energy states in the manifolds, or even the individual dipoles~\cite{SI}). In either case, a suitable combination of suppression and (passive or active) reinitialisation achieves significant steady-state population in the BTTS. To be conservative, we shall in the following primarily consider the more involved case of active reinitialisation. Naturally, this incurs a power cost to the operation, so that we will be interested in producing a total power output that is sufficiently high to cover any reinitialisation power cost and still leaves positive net produced power for light-harvesting.

To extract the energy of absorbed photons, we introduce a central trap that is equidistant from all absorbers, analogous to a photosynthetic reaction centre~\cite{Sumi2001,Law2004}. Following the quantum heat engine model~\cite{Creatore2013a, Scully2011,Dorfman2013a}, the trap is modelled as a 2LS with energy $\omega_t = \omega_{\text{good}}$, whose (incoherent) decay to the ground state at rate $\gamma_t$ represents the energy conversion process.\footnote{Incoherent decay of a single 2LS can qualitatively capture the effect of exciton transfer down a chain of sites in a tight-binding model~\cite{Giusteri:2015aa,Schaller:2016aa}.} 

Defining the steady-state population of the trap's excited and ground states as $\braket{\rho_\alpha}_{SS}$ and $\braket{\rho_\beta}_{SS}$, respectively, we assign a hypothetical current and voltage~\cite{Scully2011, Dorfman2013a}
\begin {equation}
\label{eq:qhe}
I=e\gamma_t\braket{\rho_\alpha}_{SS} ~, \quad eV= \omega_t+k_B T_{\text{vib}} \ln \bigg( \frac{\braket{\rho_\alpha}_{SS}}{\braket{\rho_\beta}_{SS}}\bigg) ~,
\end{equation}
where $k_B$ is Boltzmann's constant and $T_{\text{vib}} = 300$~K is the ambient (phonon) temperature. The second term in the voltage expression ensures thermodynamic consistency~\cite{SI}. Optimising the `load' via $\gamma_t$ yields the maximally achievable output power $P_{\text{max}} = I (\gamma_t^*) \cdot V(\gamma_t^*)$ at optimal $\gamma_t^*$. To obtain the net power of such a photocell, we must account for the energetic expenditure associated with the reinitialisation process. To this end, we apply a similar concept to the reinitialisation process: we assign a voltage and (upwards current) for each reinitialisation step and sum over the respective products to obtain the total reinitialisation power~\footnote{We have here dropped the second term in the voltage definition of Eq.~\eqref{eq:qhe}. This is a conservative choice and errs on the side of caution: it will over- rather than underestimate the input power~\cite{SI}.}.

We now proceed to solve the dynamics using a Bloch-Redfield approach~\cite{Breuer2002}: non-secular Bloch-Redfield dissipators are formed from the optical, Eq.~\eqref{eq:Hoint}, and vibrational, Eq.~\eqref{eq:Hpint}, Hamiltonians. This approach is most appropriate for relatively weak vibrational coupling, but we discuss the case strongly coupled vibrational baths in the polaron frame~\cite{Nazir_2016} in the SI~\cite{SI}. Apart from the (half-sided) photonic band-gap, we use the free-space optical spectral density, and assume radiative equilibrium with the sun, $T_{\text{opt}}=5800$~K~\cite{Wurfel:2016aa, Higgins2017}. Phonon processes are based on an Ohmic room-temperature bath, with typical rates exceeding optical ones by three orders of magnitude.
Further dissipators describe extraction, trap decay and reinitialisation processes. The chosen extraction rate $\gamma_x$ is comparable to that of typical phonon processes. The steady state of the system is then obtained as the nullspace of the total Liouvillian; full details of this process, all parameters, photon and phonon spectral densities, and the explicit master equation can be found in the SI~\cite{SI}.

\begin{figure}
\centering
	\includegraphics[width=0.47\textwidth]{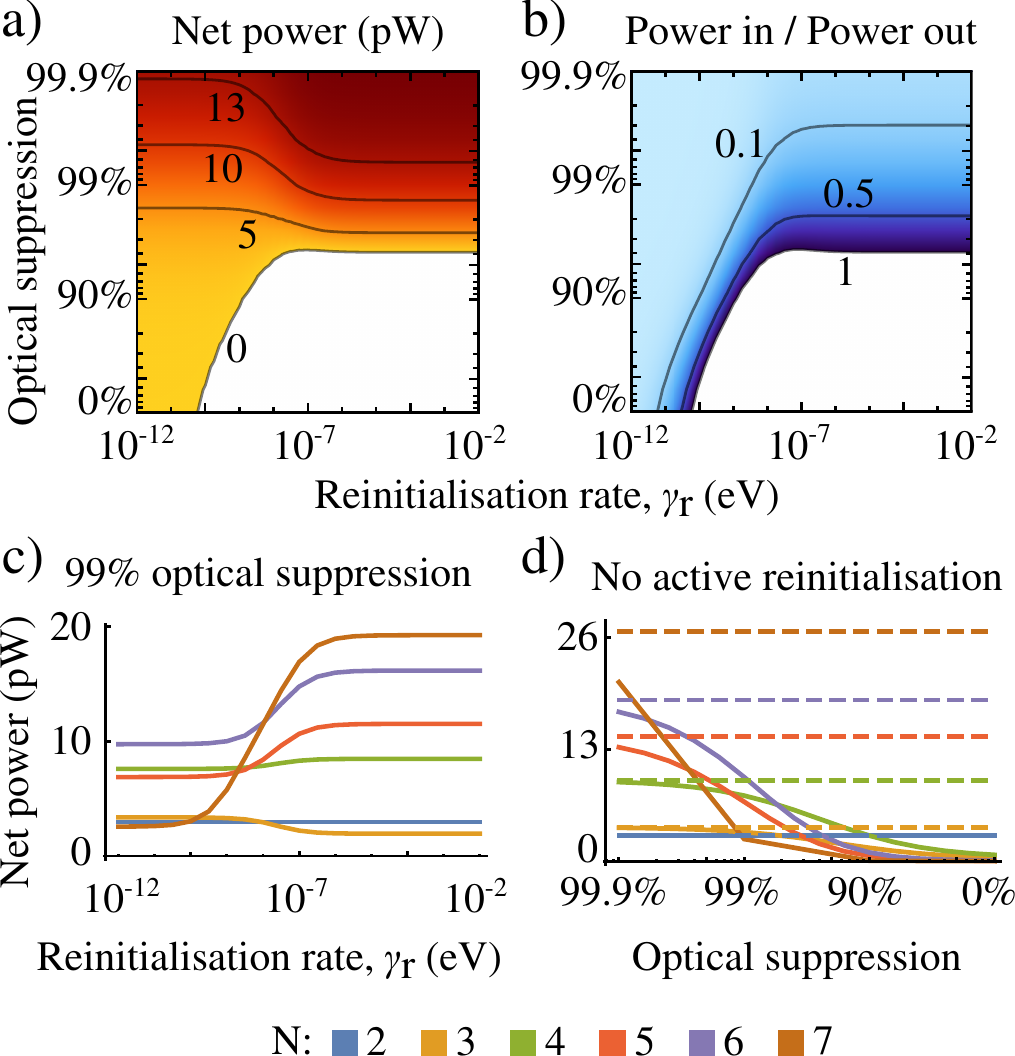}
\caption{a) Net power produced by a pentamer as a function of decay suppression and reinitialisation rate. b) Ratio of the power in against the power out across the same parameter scan. c) A cross-section at 99\% optical suppression for varying $N$. d) How different suppression strengths affect net power production without reinitialisation. Dashed lines denote 100\% suppression. Full details of the parameters used are in the SI~\cite{SI}.}
\label{fig:pscan}
\end{figure}

\textit{Results} -- First we explore the interplay between the degree of suppression and actively applied reinitialisation. As a function of these parameters, Figs.~\ref{fig:pscan}a and b show, respectively, the net power produced and the fraction of expended input vs output power, both for the case of a pentamer. For poor suppression, the system easily leaks down from the BTTS, potentially even resulting in a negative overall power balance (white region) under faster reinitialisation. In an intermediate suppression regime, active reinitialisation becomes worthwhile, so that faster reinitialisation improves performance further.  In this regime, the low fraction of output power that is re-invested into active reinitialisation (see Fig.~\ref{fig:pscan}b) indicates net power is not at risk of being easily overcome, even when a practical implementation for reinitialisation proves to be somewhat more energetically costly than assumed in our analysis. Finally, for high optical suppression the guide-slide effect successfully produces substantial (net) power in the absence of active reinitialisation, see Fig.~\ref{fig:pscan}d for an illustration of this transition into passively self-reinitialising operation. A longer discussion on how suppression enables passive `free' reinitialisation can be found in Sec.~IC of the SI \cite{SI}.

\begin{figure}
\centering
	\includegraphics[width=0.47\textwidth]{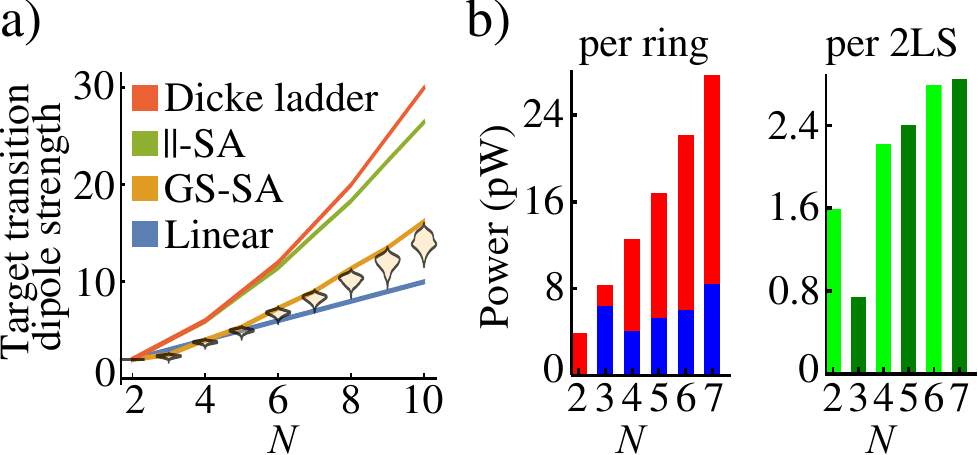}
\caption{a) Scaling of oscillator strength of the light-havesting target-transition with system size for the different scenarios discussed in the main text. 
For GS-SA we also include the beige patches corresponding to distributions for 5\% disorder introduced across model parameters (10\,000 trials for up to $N=7$, then 1\,000, 500 and 50 trials for $N=8,9,10$, respectively). b) Output (red) and input (blue) power for different system sizes with 99\% suppression of undesired optical modes, and superlinear growth of net power per site (green). Full details of the parameters used are in the SI~\cite{SI}.
}
\label{fig:superscaling}
\end{figure}

To demonstrate superabsorption, i.e.~superlinear scaling of photon absorption and energy conversion, we now investigate the effect of increasing the ring-size $N$. Fig.~\ref{fig:superscaling}a shows target transition oscillator strength as a function of $N$: we see that GS-SA displays superlinear scaling for $N>3$ but trails behind $||$-SA and the uncoupled Dicke model~\footnote{Note that the gradient changes every two data points since the target transition lies at the centre of the ladder for a ring with an odd $N$, whereas for an even-sized ring it sits just below the middle}, due to a combination of employing interacting and skewed dipoles. A similar departure from the quadratic scaling of the idealised Dicke model has also recently been observed experimentally for colour centres in diamond owing to system inhomogeneities~\cite{Angerer2018}. 
Encouragingly, our predicted superlinear GS-SA advantage persists upon introducing 5\% normally-distributed disorder amongst relevant parameters (energy splittings $\omega_A$, natural lifetimes $\tau_L$, positions $\vec{r}_i$, and dipole orientations), as shown by the pale distribution marks. Further information regarding disorder can be found in the SI~\cite{SI}.

Focusing on the generated net power, Fig.~\ref{fig:pscan}c shows the transition into the guide-slide regime for $2 \leq N \leq 7$ as a function of $\gamma_r$ as active reinitialisation is implemented for fixed suppression at 99\%. 
Once $\gamma_r$ becomes large enough to compensate for (suppressed) optical leakage from the BTTS, larger rings do indeed achieve higher net power. This superlinear growth of the net power with $N$ is confirmed by Fig.~\ref{fig:superscaling}b, which also gives a breakdown of input vs output power. We note that the dimer and trimer are special: the former performs well as it does not require reinitialisation, whereas the latter has insufficient collective-enhancement for making active reinitialisation worthwhile (but performs adequately left to its own devices, see SI~\cite{SI}). Beyond $N>3$, however, there is an increasing trend of GS-SA enabling quantum-enhanced photocell performance.

\begin{figure}
\centering
	\includegraphics[width=0.47\textwidth]{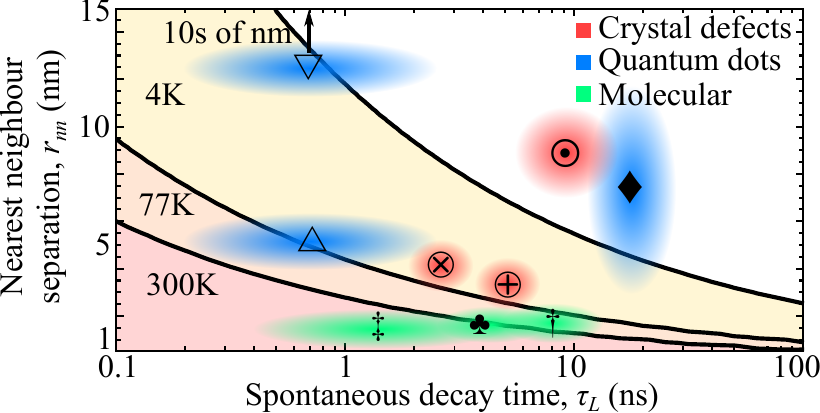}
\caption{Map of where a quadmer produces positive net power as a function of 2LS natural lifetime and nearest neighbour separation. Regions typical of potential candidate systems are overlaid: Adjacent ($\bigtriangledown$) and stacked $\bigtriangleup$) self-assembled InGaS quantum dots~\cite{Creasey2012a,Gerardot2005a}. $\diamondsuit$: Colloidal quantum dots~\cite{Lounis2000,Brichkin2016}. $\odot$: Nitrogen vacancies~\cite{Albrecht2014,Neumann2010a,Angerer2018}. $\otimes$: Silicon vacancy centres~\cite{Rogers2014a,Sipahigil2016,Morse2017}. $\oplus$: Phosphorous defects~\cite{OBrien2001,Fuechsle2012,Schofield2003}.$\clubsuit$: BODIPY dye~\cite{Berezin2010,Pochorovski2014,Acikgoz2010,Chung2016}. $\dagger$: Porphyrin rings~\cite{OSullivan2011a,Sprafke2011,JunhuaYu2004}. $\ddagger$: Merocyanine dye (H-aggregate) and pseudoisocyanine chloride (J-aggregate)~\cite{Rosch2006,Wurthner2016,Obara2012,VonBerlepsch2000}. 
} 
\label{fig:results}
\end{figure}

Focusing on the example of a quadmer we explore the temperature dependence of the effect. As a function of the 2LS spontaneous emission time $\tau_L$ and nearest neighbour separation $\vert \vec{r}_{i,i+1} \vert$ -- the key parameters which determine the oscillator and F\"orster coupling strengths -- we map out the region achieving net positive power for different phonon temperatures in Fig.~\ref{fig:results}. For this purpose, the dipole orientation angles were optimised, but never departed far from our previously employed (slightly sub-optimal) ad-hoc choice of $\theta_{\text{eq}} = \pi/2, \theta_{\text{zen}} = \pi/4$ (see Fig.~S3 \cite{SI} for the typical dependence of net power on the dipolar offset angles). Fig.~\ref{fig:results} shows that cooling below room temperature expands the working range of GS-SA, since a colder phonon environment boosts the directionality of `guide-sliding' onto ladder states even for closely spaced levels within the manifold (i.e.~weaker F\"orster couplings).

To address the question of potential candidate systems suitable for exploring the GS-SA effect, we indicate regions and data-points referring to the properties of several state-of-the-art platforms for nanostructure photonics. This demonstrates a broad range of credible building blocks for GS-SA antennae, leaving the key challenges of (i) ring assembly with control over the direction of the optical dipoles,\footnote{In crystal defects, possible dipole angles are set by the axis of the crystal~\cite{Doherty2013}, potentially restricting possible configurations which could be synthesised in a single bulk crystal.} (ii) embedding into a suitably engineered photonic environment, and (iii) furnishing the antennae with efficient energy extraction~\cite{Dubi2015,Baghbanzadeh2016,Zhang2017} and reinitialisation channels. Finally, of particular relevance to molecular systems, exciton-exciton annihilation needs to be controlled or avoided~\cite{Higgins2017}. The possibility of constructing molecular light-harvesting ring systems of dyes has already been considered using DNA origami~\cite{Hemmig2016}, though the separations are currently larger than desired for room temperature GS-SA applications. A detailed discussion of the effect of non-radiative decay processes in these systems can be found in the SI~\cite{SI}.

\textit{Summary} -- We have proposed a set of intuitive requirements for guide-slide superabsorption. We have shown that frequency-selective passive reinitialisation via photonic band-gap suppression and, if required, additional active reinitialisation results in a superabsorbing steady-state, characterised by a superlinear scaling of optical absorption and net power conversion with increasing system size. Inspired by photosynthetic structures, we have presented an example ring-system exhibiting this effect. Our proposed setup not only remains viable in the presence of phonons, but even benefits from vibrational relaxation. Further, the effect proves remarkably robust to substantial amounts of disorder and realistic imperfections, showing promise for experimental exploration across a number of platforms. Importantly, bio-inspired molecular rings, similar to the ones described in Refs.~\cite{OSullivan2011a,Sprafke2011}, could function at room-temperature, offering an exciting perspective for nanophotonics, quantum-enhanced light-harvesting, and possibly future approaches to organic photovoltaics. 

\acknowledgements

We thank Dale Scerri for useful discussions. 
W.M.B. acknowledges studentship funding from EPSRC under grant no.~EP/L015110/1.
E.M.G. thanks the Royal Society of Edinburgh and the Scottish Government for support.

%

\pagebreak
\begin{center}
\textbf{\large Supplementary Information: Light-harvesting with guide-slide superabsorbing condensed-matter nanostructures}
\end{center}
\setcounter{equation}{0}
\setcounter{figure}{0}
\setcounter{table}{0}
\setcounter{page}{1}
\makeatletter
\renewcommand{\theequation}{S\arabic{equation}}
\renewcommand{\thefigure}{S\arabic{figure}}
\renewcommand{\bibnumfmt}[1]{[S#1]}
\renewcommand{\citenumfont}[1]{S#1}

\newcommand\varmp{\mathbin{\vcenter{\hbox{%
  \oalign{\hfil$\scriptstyle-$\hfil\cr
          \noalign{\kern-.5ex}
          $\scriptscriptstyle({+})$\cr}%
}}}}


\section{Introduction}

This supporting documentation contains full details of the model used in the main text, including tables listing a full set parameters for all figures. In particular, it includes more detail on the light-matter and inter-dipole coupling for non-parallel optical dipoles, the construction of the optical and vibrational Bloch-Redfield tensors, and our method for finding steady state solutions. Further, we explicitly demonstrate the breakdown of $||$-SA and the robustness of GS-SA in the presence of phonons, as well as confirming the integrity of superlinear GS-SA scaling for a range of extensions and variations of the model from the main text. We also discuss the effect of non-radiative relaxation, and potential pathways for mitigating its detrimental effect on the output power. Finally, we give a detailed discussion of the disorder applied to the GS-SA model, finding that it takes disorder of order 10\% or more to break the principles enabling GS-SA introduced in the main text.

\section{Optical control and coupling}

\subsection{GS-SA complete optical process map}

\begin{figure}
\centering
	\includegraphics[width=0.45\textwidth]{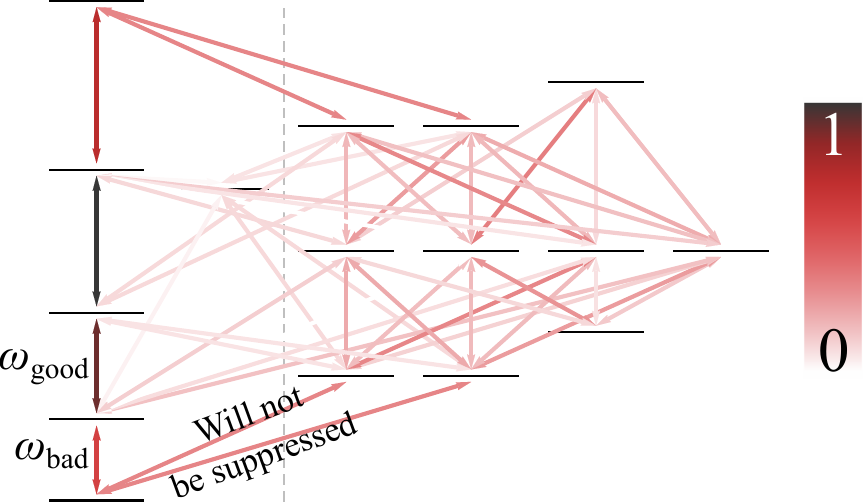}
\caption{The optical process map for a GS-SA quadmer with the same parameters as those used in Fig.~2 of the main text. All processes arising from the coupling Hamiltonian (and included in our numerical model) are shown. Notably, several processes now cross the dashed dividing line. The colouring of each line denotes the relative strength of each process, normalised against the strongest process on the map. For clarity, energy separations between the different states in this image are not to scale: intra-band spacings have been scaled up with respect to interband spacings.}
\label{fig:3Doptical}
\end{figure}

The optical process map for GS-SA presented in Fig.~2 of the main text only accounts for the collective coupling (in the direction perpendicular to the plane of the ring), $\vec{D} = \sum_i^N \vec{d}_i$, which 71\% of the overall dipole moment aligns with. The collective enhancement which we are primarily interested in can be sufficiently explained using just this shared direction, but our model also accounts for coupling in the plane of the ring. The overlap of optical dipole directions in the plane of the ring means that for GS-SA one no longer has the total optical separation of symmetric and non-symmetric states that is present in $||$-SA (denoted by the vertical dashed line in the figure). A complete process map of all optical processes is given in Fig.~\ref{fig:3Doptical}.

\subsection{Photonic band-gap suppression}

To apply optical suppression for GS-SA, we define a cut-off frequency as the midpoint between the transition frequency for the target transition, $\omega_\text{good}$, and the frequency associated with the transition below the bottom of target transition state (BTTS) down to the rung beneath, $\omega_\text{bad}$. Any optical process with an associated transition frequency which is smaller than the cut-off will be suppressed. Naturally, this suppression affects both the up and down rates of the processes.

\begin{figure}
\centering
	\includegraphics[width=0.45\textwidth]{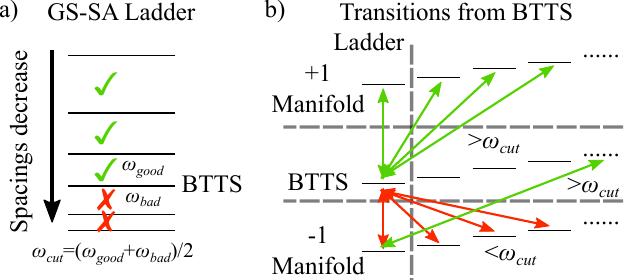}
\caption{a) Demonstration of how the displaced ladder states in GS-SA are used to define a cut-off frequency, $\omega_{cut}$. b) Schematic showing how our choice of $\omega_{cut}$ will cover all potential optical relaxation frequencies from BTTS, while still keeping relaxation pathways. An example `free reinitialisation' pathway is also included. For clarity, energy separations between the different states in this image are not to scale: intra-band spacings have been scaled up with respect to interband spacings.}
\label{fig:GoodBadCartoon}
\end{figure}

This cut-off frequency definition implies that all the collectively enhanced optical transitions will be suppressed beneath the target transition, as well as relaxation to any other (off-ladder) state in the manifold below. This relies on the fact that in GS-SA, the non-degenerate, enhanced optical transition frequencies increase as one moves up the ladder, as can be seen by the spacings of the states in Fig.~2 of the main text, as well as Figs.~\ref{fig:3Doptical} and \ref{fig:GoodBadCartoon}. For the case of $||$-SA, a similar process can be applied if instead one suppresses all modes {\it above} a certain threshold rather than below. 

In reality, photonic band-gaps are of finite width (and often not of perfect square well form) as opposed to a binary partition of frequency space. A wide enough band-gap may nevertheless suppress all ladder transitions below the BTTS. However, whether or not that is the case, we find our approach works well as long as emission at frequency $\omega_\text{bad}$ and it its immediate vicinity is adequately suppressed.

We note that alternatively, one could seek to selectively enhance the target transition~\cite{Higgins2014aSUP} (or indeed a range of transitions above a threshold frequency). Modifying exciton dynamics of photosynthetic biological ring antennae with the help of structured photonic environments is the subject of this recent study~\cite{Saez-Blazquez2019SUP}.

\subsection{``Free" ratcheting reinitialisation}

Analysing the complete optical process map in Fig.~\ref{fig:3Doptical}, we see that for GS-SA there are additional optical pathways. In the main text we focussed on discussing only on the collectively-enhanced optical transitions, linking states which are optically disconnected from `off-ladder' system eigenstates (whilst all pathways were included in the calculations). For GS-SA, however, each (bottom of excitation manifold) ladder state is optically connected to the other states in adjacent manifolds above as well as below. 

As mentioned above, beneath the target transition in GS-SA the suppression covers all transitions from the ladder states to the entirety of the manifold below, and is thus effective for all possible decay pathways. By contrast, processes linking ladder states to higher energy off-ladder states in the manifold above are not necessarily suppressed: these transitions can potentially have a frequency larger than the cut-off, meaning optical excitation to the manifold above remains possible (see Fig.~\ref{fig:3Doptical} and Fig.~\ref{fig:GoodBadCartoon} for an example of such a transition). If such an excitation occurs, then rapid phonon relaxation takes the system down to the bottom of the new manifold, from where optical relaxation is suppressed, but excitation is once more still possible. This mechanism allows for the system to climb the ladder without external reinitialisation. The quality of this self ladder climbing depends on the strength of the suppression. 

 In the main text, Fig.~3d shows this effect: in the absence of any active reinitialisation, the net power produced increases as suppression strength increases. This is true for any ring size, but larger rings require a higher degree of suppression to fully passively self-reinitialise. Conversely, Fig.~3c demonstrates that for a fixed degree of optical suppression at 99\% and for rings larger than $N=3$, additional active reinitialisation will lead to a substantial (superlinear) increase of output power.  Noteworthy cases are the dimer (which does not require any active reinitialisation) and the trimer which intriguingly performs worse under active reinitialisation: this is because here the cost of having fast external reinitialisation to steer more of the system into the superabsorbing regime is not worth the gains in power produced, owing to the limited degree of  collective enhancement of a small trimer system. Whilst larger rings clearly benefit from active reinitialisation at 99\% suppression, it is interesting that free reinitialisation also works reasonably well for climbing up to two manifolds (i.e. up to the BTTS of a hexamer). As mentioned, larger rings featuring three or more manifolds beneath the target extraction transition require a higher degree of suppression for passive reinitialisation to be viable (c.f. Fig.~3d).

This mechanism of free reinitialisation is reminiscent of the optical ratcheting effect proposed by Ref.~\cite{Higgins2017SUP}, however with an inverted energy landscape in each excitation manifold due to the change of sign of dipole-dipole couplings (in molecular language effectively corresponding to a switch from H- to J-aggregate character). A further difference is that the present effect requires a photonic band-gap in the current setup, as opposed to Ref.~\cite{Higgins2017SUP}.

\section{Optical and Vibrational Dissipators}
\label{sec:diss}

Our approach for dealing with the optical and vibrational environments closely follows the methods outlined in Ref.~\cite{Breuer2002SUP}.

\subsection{Electromagnetic environment}
\label{sec:dissOPT}

As a starting point for dealing with the optical environment, we take our optical interaction Hamiltonian,
\begin{equation}
\hat{H}_{I,\text{opt}}=\sum_{i=1}^N \vec{d}_i\hat{\sigma}_i^x \otimes \sum_k f_k (\hat{a}_k+\hat{a}_k^\dagger)~,
\end{equation}
and rewrite the individual dipoles' $\hat{\sigma}^x$ operators in terms of system eigenbasis states [using the matrix diagonalising the Hamiltonian Eq.~(1) of the main text]. This yields the optical transitions between the system eigenstates. Reflecting the vectorial nature of the optical dipoles, each term is a 3-vector, with one element per Cartesian direction of the optical field. It is easy to see that optical transitions will only link eigenstates from different (adjacent) excitation manifolds and cannot give rise to intra-manifold transitions.

The system operators (in the eigenbasis) of the light matter interaction can be written as a Hermitian matrix with each element being a 3-vector. Terms beneath the diagonal represent processes which lower the system energy whereas those above raise it. By taking a list of all processes, i.e.~all non-zero vectors in this matrix, we form the (non-secular) Bloch-Redfield tensor (in the Schr\"{o}dinger picture):
\begin{multline}
\mathcal{D}_\text{opt}=
\sum_{n,m}\vec{d}_n\cdot\vec{d}_m\Big(A_m(\omega_m)\rho_S(t)A^\dagger_n(\omega_n)\Gamma_{nm}(\omega_m) \\
+ A_n(\omega_n)\rho_S(t)A^\dagger_m(\omega_m) \Gamma^\dagger_{nm}(\omega_m) \\
- \rho_S(t)A^\dagger_m(\omega_m)A_n(\omega_n)\Gamma^\dagger_{nm}(\omega_m) \\
- A^\dagger_n(\omega_n)A_m(\omega_m)\rho_S(t) \Gamma_{nm}(\omega_m)\Big)~.
\end{multline}
This includes all pairwise combinations of lowering (raising) processes $A_\alpha^{(\dagger)}$ with associated frequency $\varmp\omega_\alpha$.
Every term in the above expression is weighted by the dot product of the two 3-vectors associated with the processes being paired. Note that here the $\vec{d}_{n,m}$ no longer represent the dipole moments of individual dipoles but those associated with system energy eigenstates. 

The above dissipator includes environmental correlation functions, evaluation of which yields the process rates. The correlation functions are
\begin{align}
\Gamma_{nm}(\omega) &= \int_0^\infty ds e^{i\omega s} \Braket{B^\dagger_n(t) B_m(t-s)} ~, \nonumber \\
\Gamma^\dagger_{nm}(\omega) &= \int_0^\infty ds e^{-i\omega s} \Braket{B^\dagger_m(t-s) B_n(t)}~,
\end{align}
where the $B(t)$ are the environment operators of the interaction Hamiltonian. Since all dipoles are assumed to be co-located at the same position (for the present purpose of deriving dissipative processes~\footnote{This is a good approximation as long as the separations between dipoles are much smaller than the wavelength of relevant photons. } -- we do allow for separation when calculating dipole-dipole interactions), there are no phase factors in coupling elements and we can write a single joint correlation function of the general form~\footnote{It should be noted that the $\delta_{ij}$ in the correlation function is responsible for the dot product of the dipoles in the optical dissipator: it ensures that there is no contribution for pairs of terms pointing along orthogonal Cartesian directions.}
\begin{equation}
\Gamma_{n_im_j}(\omega) = \delta_{ij}\bigg(\frac{1}{2}\gamma(\omega)+iS(\omega)\bigg)~,
\end{equation}
where
\begin{equation}
\gamma_{\text{opt}} (\omega)=\kappa_{\text{opt}}\omega^3\big(1+n(\omega)\big) ~.
\end{equation}
Here, $\kappa_{\text{opt}}$ is determined by the lifetime $\tau_L$ of an isolated 2LS with splitting $\omega_A$, such that $\gamma_{\text{opt}}(\omega_A) = \tau_L^{-1}$ at zero temperature [so that $n(\omega) = 0$]. The cubic frequency dependence of $\gamma_{\text{opt}}(\omega)$ arises from the density of optical modes in free space, and the Bose-Einstein occupancy of photon modes $n(\omega)$ is
\begin{equation}
n(\omega)=\frac{1}{e^{\beta\hbar\omega}-1} ~
\end{equation}
with $\beta = 1 / k_B T$, where $T$ is the photon bath temperature and $k_B$ is Boltzmann's constant.

For the purpose of light harvesting, we set $T=5800$~K matching the solar temperature. This choice corresponds to being in radiative equilibrium with a black-body emitter at the temperature of the sun; the Supplement of Ref.~\cite{Higgins2017SUP} contains a more detailed discussion of the implications of this assumption.

The correlation function's imaginary contribution $S(\omega)$ results in unitary corrections to the Hamiltonian dynamics through generalised Lamb and Stark shift terms. The most important consequence of those is the cross Lamb-shift dipole-dipole interaction~\cite{Higgins2014aSUP}, which we have already included explicitly in our ring Hamiltonian Eq.~(1). By contrast, diagonal Lamb shift terms only slightly renormalise the transition energy itself and can be absorbed in the effect splitting of the dipoles. For these reasons, we here neglect $S(\omega)$.

Finally, all rates with an associated frequency below the suppression threshold have the photonic band-gap suppression applied, by multiplication with the respective suppression factor.

\subsection{Vibrational}

The steps for finding the vibrational dissipator are similar to those taken for the optical one. We start with the spin-boson interaction Hamiltonian repeated across the $N$ sites (labelled with index $i$):
\begin{equation}
\hat{H}_{I,\text{vib}}=\sum_{i=1}^N \hat{\sigma}_i^z \otimes \sum_q g_{i,q} (\hat{b}_{i,q}+\hat{b}_{i,q}^\dagger) ~.
\end{equation}
An important difference to the optical case is that we now assume an independent phonon bath for each absorber. Hence, there will be no collective effects caused by phonons. Rather, we treat each phonon bath separately (all being equivalent and thus identical copies of each other).

We now have a system interaction matrix for each system, which upon transformation to the system eigenbasis result in $N$ matrices of raising and lowering processes between system eigenstates. For each matrix, $\alpha$, a list of all the processes is taken to form another non-secular Bloch-Redfield dissipative term:
\begin{multline}
\mathcal{D}_{\alpha,\text{vib}}=
\sum_{n,m}w_nw_m\Big(A_m(\omega_m)\rho_S(t)A^\dagger_n(\omega_n)\Gamma_{nm}(\omega_m) \\
+ A_n(\omega_n)\rho_S(t)A^\dagger_m(\omega_m) \Gamma^\dagger_{nm}(\omega_m) \\
- \rho_S(t)A^\dagger_m(\omega_m)A_n(\omega_n)\Gamma^\dagger_{nm}(\omega_m) \\
- A^\dagger_n(\omega_n)A_m(\omega_m)\rho_S(t) \Gamma_{nm}(\omega_m)\Big) ~,
\end{multline}
where the $n,m$ summation covers all pairwise combinations of processes for that matrix, and the factors $w_i$ reflect appropriate weightings arising from the transformation to the ring eigenbasis. 
The full vibrational dissipator is then given by the sum over $\alpha$,
\begin{equation}
\mathcal{D}_{\text{vib}}=\sum_\alpha^N\mathcal{D}_{\alpha,\text{vib}} ~.
\end{equation}
Since the environments for each set of terms can be considered identically, there is only one form of correlation function to calculate:
\begin{equation}
\Gamma_{n_im_j}(\omega) = \delta_{ij}\bigg(\frac{1}{2}\gamma(\omega)+iS(\omega)\bigg) ~.
\end{equation}
Assuming a structureless Ohmic spectral density (see Sec.~\ref{sec:specdens} for other choices), we write
\begin{equation}
\gamma_{\text{vib}}(\omega)=\kappa_{\text{vib}}\omega\big(1+n(\omega)\big) ~,
\end{equation}
where for now we let $\kappa_{\text{vib}}$ be a phenomenologically motivated prefactor which sets the timescale of phonon rates. Reflecting the fact the vibrational processes typically occur on a much faster timescale than optical ones, we define $\kappa_{\text{vib}} = 10^3 \times \gamma_{\text{opt}}(\omega_A) / \overline{ \omega_{\text{vib}}}$, where $\overline{ \omega_{\text{vib}}}$ represents the mean intra-excitation-manifold transition frequency. This ensures that (spontaneous) phonon emission at the average phonon energy proceeds with a rate that is 1000 times faster than the spontaneous photon emission time of an individual optical dipole.

The Bose-Einstein factor $n (\omega )$ is defined as in the optical case but now for the temperature of the phonon bath at 300~K (except for Fig.~5 in the main text which explores different temperatures). As is the case with the optical dissipator, the imaginary component of the environment correlation functions is discarded.

\section{Calculating the power output}

\subsection{Discussion of quantum heat engine approach}

The quantum heat engine approach~\cite{Scully:2010aaSUP,Dorfman2013aSUP} is based on the idea of introducing an abstract two-level system `load' (hereafter referred to as the `trap') to a photocell as a means of obtaining a relationship between current and voltage. The `resistance' of the load is tuneable by changing the phenomenological decay rate $\gamma_t$ from the higher energy state $\vert \alpha \rangle$ into the lower state $\vert \beta \rangle$. Scanning $\gamma_t$ (typically through many orders of magnitude), one obtains an $IV$-curve, and via $P=IV$ also the power as a function of voltage.

Specifically, the current is given by
\begin {equation}
I=e\gamma_t \braket{\rho_\alpha}_{SS} ~,
\label{eq:current}
\end{equation}
where $\braket{\rho_i}_{SS}$ denotes the steady state population of the state $i$, and the effective potential difference is
\begin{equation}
e V = \hbar \omega_{\alpha\beta}+k_B T_{\text{vib}} \ln \bigg( \frac{\braket{\rho_\alpha}_{SS}}{\braket{\rho_\beta}_{SS}}\bigg) ~,
\label{eq:voltage}
\end{equation}
where $\hbar\omega_{\alpha\beta}$ is the energy of the trap, $k_B$ is Boltzmann's constant, and $T_{\text{vib}}$ is the ambient (phonon bath) temperature. The logarithmic term in the potential difference equation provides a correction to the effective voltage which is based on the deviation of the trap populations from the thermal distribution~\cite{Creatore2013aSUP}. Its inclusion ensures thermodynamic consistency: typically, the effective voltage $V$ will be lower than $\hbar\omega_{\alpha\beta} / e$. This implicitly partitions the trap decay process into a heat and a work current contribution. By contrast, neglecting this partitioning may result in the thermodynamic inconsistencies described by Ref.~\cite{Gelbwaser-Klimovsky:2017aaSUP}.

We employ this model for obtaining the generated power (`Power out') of our photocell devices. All results shown are obtained for a numerically optimised $\gamma_t^*$ that delivers peak power at $\max (I(\gamma_t^*) \cdot V(\gamma_t^*))$.

For reinitialisation (`Power in'), we distinguish three cases: (i) `ladder-climbing' reinitialisation (ii) `lossy ladder-climbing' reinitialisation and (iii) site-based reinitialisation. For the latter case -- employed in Secs.~\ref{sec:siteinit} and \ref{sec:cohexsiteinit} of this document -- we simply employ the same process in reverse, but this time separately for {\it each dipole}. Here, the roles of `upper' and `lower' level are swapped in Eqs.~\eqref{eq:current} and \eqref{eq:voltage} to the dipoles' ground and excited states, respectively. Additionally, $\gamma_t$ in Eq.~\eqref{eq:current} is replaced by the applied reinitialisation rate $\gamma_r$. The total input power is then given by the sum over the power terms associated with each dipole $P_{\text{in}} = \sum_{i = 1}^{N} I_i \cdot V_i = N I_0 V_0$ (the second equality follows in the absence of disorder when all dipoles are equivalent). Note that this particular type of reinitialisation indiscriminately moves the system up the next higher exciton manifolds, with no preference for populating ladder states. This is not a problem for GS-SA, where phonon processes will let it slide back onto the ladder, but precludes its use for $||$-SA.

For ladder-climbing reinitialisation -- considered in the main paper -- as well as lossy ladder climbing reinitialistion -- we instead calculate $P_{\text{in}} = \sum_{n = 1}^{\lfloor (N-1)/2 \rfloor} I_n \cdot V_n$, where $I_n$ is the upwards current of ladder rung $n$ and $V_n$ the associated voltage. Reinitialisation only takes place between the ground state and the BTTS (situated at rung $\lfloor (N-1)/2 \rfloor$ counted from the bottom of the ladder). In this case the logarithmic term in Eq.~\eqref{eq:voltage} is questionable, so we directly identify the respective rung transition frequency as $V_n$. However, we have checked that inclusion of the logarithmic term over the range of applied of $\gamma_r$ only makes a negligible difference to the results (and leads to a marginally reduced input power). This not only confirms that our choice is conservative but further suggests that a thermodynamic correction is not required for our purpose.

\subsection{Quantum heat engine processes}

Implementing the above-described quantum heat engine approach requires additional dissipators to be added to the master equation describing the dynamics of the ring antenna (as well as an extended Hilbert space dimension to accommodate the additional trap two level system).

These can be introduced through operationally motivated Lindblad dissipators of the generic form~\cite{Breuer2002SUP}
\begin{equation}
\mathcal{D}_L[\rho, \gamma, \hat{L}] \equiv \gamma \left( \hat{L} \rho \hat{L}^\dag - \frac{1}{2} \{ \hat{L}^\dag \hat{L}, \rho \} \right) ~,
\end{equation}
where $\{ \bullet , \star \} = \bullet \star + \star \bullet$ denotes the anti-commutator, $\hat{L}$ is a Lindblad operator and $\gamma$ its associated rate. However, in keeping with treating the optical and vibrational environment to Bloch-Redfield level of theory, we shall employ that approach for these dissipative terms as well, noting that in certain cases our approach automatically reduces to the simpler Lindblad form.

To construct the dissipator of incoherent processes, we start with a normalised interaction Hamiltonian (matrix) $M$, which features non-zero matrix transition elements connecting the desired levels. Upon transformation of this matrix into the system's eigenbasis, we obtain the corresponding Bloch-Redfield dissipator by summing over all pairwise combinations of the non-zero interaction elements, yielding:
 \begin{multline}
\mathcal{D}_{BR} = \gamma ~
\sum_{n,m} \Theta(\sigma \omega_m) \Big(A_m(\omega_m)\rho_S(t)A^\dagger_n(\omega_n) \\
+ A_n(\omega_n)\rho_S(t)A^\dagger_m(\omega_m) 
- \rho_S(t)A^\dagger_m(\omega_m)A_n(\omega_n) \\
- A^\dagger_n(\omega_n)A_m(\omega_m)\rho_S(t)\Big).
\label{eq:BRdiss}
\end{multline}
Here, $\gamma$ is the phenomenological rate of the process and $\Theta(\omega)$ is the Heaviside function (0 for $\omega < 0$ and 1 for $\omega \geq 0$). For $\sigma = 1$ this ensures that only `downwards' (decay) processes survive, whereas for $\sigma = -1$ it is uni-directional in the `upwards' direction (as will be required for reinitialisation pumping). It should be noted that since all these processes are one-way, we only take pairwise combinations of terms which both either raise or lower the system energy, creating a partial secular approximation.

\subsubsection{Trap decay operator}

As discussed above, the quantum heat engine `trap' is a two-level system which undergoes incoherent decay from its excited state $\ket{\alpha}$ to its ground state $\ket{\beta}$. This is accomplished by feeding the interaction matrix $M = \hat{\sigma}_t^- + \hat{\sigma}_t^+$ [with $\hat{\sigma}_t^- \equiv \ket{\beta}\bra{\alpha} $ and $\hat{\sigma}_t^+ = (\hat{\sigma}_t^-)^\dag$] into Eq.~\eqref{eq:BRdiss}. Further, we set $\sigma = 1$ (so that only the decay terms are effective) and choose $\gamma_t$ as the decay rate.

For the case of incoherent extraction (see below), the above procedure results in the effective Lindblad dissipator $\mathcal{D}_{t} := \mathcal{D}_L[\rho_S, \gamma_t, \hat{\sigma}_t^-]$, whilst it retains a more general form for coherent extraction, when trap and ring eigenstates become intertwined.

Going beyond the canonical heat engine picture and with reference to the extraction chain depicted in Fig.~1 of the main text, we note that in the right limit, incoherent decay of a single 2LS can also qualitatively captures the effect of exciton transfer down a chain of sites in a tight-binding model~\cite{Giusteri:2015aaSUP,Schaller:2016aaSUP}.

\subsubsection{Extraction operator}

The extraction process transfers the excitation from the ring onto the trap. In the following we describe two ways of accomplishing this.

\textbf{Incoherent:} here we consider targeted extraction, which lowers the ring population from the eigenstate at the `top-of-the-target-transition' (TTTS) to the BTTS, while at the same time exciting the trap from its ground state to its excited state. As we let the energy of the trap $\omega_t$ match the transition frequency of the target transition $\omega_\text{good}$ this process is energy conserving. Since this process is defined directly on eigenstates of ring and trap, there is no need to consider Eq.~\eqref{eq:BRdiss} and we can directly write the Lindbladian dissipator $\mathcal{D}_{x} =\mathcal{D}_L[\rho_S, \hat{L}_{\text{ex}}, \gamma_x]$ with $\hat{L}_{\text{ex}}=\ket{\text{BTTS}}\bra{\text{TTTS}} \otimes \hat{\sigma}_t^+$.

This type of extraction underlies all results of main text and also the majority GS-SA (and $||$-SA) results in this document, unless explicitly stated otherwise.

\textbf{Coherent:} Following the Supplement of Ref.~\cite{Higgins2014aSUP} we also investigate the effect of coherent Hamiltonian coupling between the ring and the trap, as an alternative model. This does not alter our core conclusions but requires a number adjustments of the model, so a full discussion of coherent extraction is given in its own section later in this document.

Physically, coherent extraction would be most naturally implemented via the same dipolar interaction between optical dipoles which gives rise to interactions between ring sites, i.e. by adding pairwise F\"orster coupling terms to the Hamiltonian (which is our approach in Sec.~\ref{sec:cohext}). Incoherent extraction could arise from the appropriate F\"orster-Dexter limit under strong vibrational coupling of both ring and trap site when the dipolar coupling is relatively small, e.g. through a more distant trap that effectively couples to ring eigenstates.

\subsubsection{Reinitialisation operators}
\label{sec:ladderR}

\textbf{Ladder-climbing:} for this reinitialisation method, we apply pumping to the BTTS from all lower energy ladder rungs. The corresponding Lindblad operators have the form $\hat{M}_{n, \text{reinit}}=\ket{\text{BTTS}}\bra{\text{ladder state:}~n}$ with $n \in \{1, \lfloor (N-1)/2 \rfloor \}$ indicating the rung from which to climb. In the absence of coherent extraction, the resulting dissipator is $\mathcal{D}_{\text{r}} = \sum_n \mathcal{D}_L[\rho_S, \hat{M}_{n, \text{reinit}}, \gamma_r]$. In the presence of coherent extraction, a dissipator based on Eq.~\eqref{eq:BRdiss} is formed for each $n$ separately with interaction matrices $\hat{M}_{n, \text{reinit}} + \hat{M}_{n, \text{reinit}}^\dag$, rate $\gamma_r$, and $\sigma = -1$. The sum over these is then the combined reinitialisation dissipator $\mathcal{D}_{\text{r}}$.

\textbf{Lossy ladder-climbing:} this reinitialisation method is essentially identical to the one above, but with the lowest energy eigenstate of the each rung being pumped to the highest energy eigenstate in the rung above. We compare these two variants of ladder climbing reinitialisation in Sec.~\ref{sec:lossylad}.

Physically, ladder climbing could be achieved via optical pumping, temporarily and periodically reversing the role of the extraction process, or by duplicating the extraction infrastructure with a dedicated second injection process. Inevitably, the details of implementing this will need to be tailored to what is feasible given a specific candidate system and architecture.

\textbf{Site-based:} for site-based reinitialisation we have $N$ interaction matrices, $M_{\alpha,\text{reinit}}=\hat{\sigma}_i^+ + \hat{\sigma}_i^-$, where $i \in \{1\dots N\}$ indexes the ring sites. Each interaction matrix results in a dissipator term via Eq.~\eqref{eq:BRdiss} with $\sigma = -1$ and at rate $\gamma_r$. The sum over all these constitutes the total $\mathcal{D}_{\text{r}}$.

Physically, this process could for example represent the injection of electron and hole charge carrier pairs on the semiconductor platform, similar to the way an LED or laser diode is powered.

\section{Finding the steady state with the Liouvillian}

The overall master equation governing the dynamics of our ring antenna plus trap is ($\hbar = 1$)
\begin{equation}
\frac{d}{dt}\rho_S= -i [\hat{H}_S,\rho_S]+\mathcal{D}_{\text{opt}}+\mathcal{D}_{\text{vib}}+\mathcal{D}_{x}+\mathcal{D}_{r}+\mathcal{D}_{t}
\label{eq:fullme}
\end{equation} 
and can be recast into the Liouvillian form
\begin{equation}
\frac{d}{dt}\rho_S = \mathcal{L}\rho_S 
\end{equation}
with the formal solution
\begin{equation}
\rho_S(t)=e^{\mathcal{L}t}\rho_S(0) ~.
\end{equation}
Diagonalising the Liouvillian, the master equation reads
\begin{equation}
\rho_S(t)=
\begin{pmatrix}
e^{\lambda_{1}t} & 0 & \cdots \\
0 & e^{\lambda_{2}t} & \cdots \\
\vdots & \vdots & \ddots 
\end{pmatrix}
\rho_S(0) ~,
\end{equation}
where $\lambda_{i}$ are the complex eigenvalues of $\mathcal{L}$. Since in most cases all the eigenstates of our system are connected, there is a unique steady state, and thus only one zero-valued eigenvalue exists (all others having negative real parts). For $t\rightarrow\infty$ only the eigenstate with associated $\Re(\lambda) = 0$ survives. The steady state of the system is then given by this eigenvector, regardless of the initial system state. 

For $||$-SA without phonons and disorder (covered in Sec.~\ref{sec:parsa}), the system eigenstates are not fully connected leading to multiple zero eigenvalues of the Liouvillian. In this case, obtaining the steady state is slightly more complicated and depends on the choice of initial state, which is the overall ground state in our case. 

\section{Full set of model parameters}
\label{sec:parameters}

A complete set of model parameters used for the figures of the main text and this documents are printed in Tab.~1 and Tab.~2, respectively. Not shown is the trap decay rate $\gamma_t$ as this is always numerically scanned over to find the value maximising the output power. For disorder calculations, the parameters are given by Gaussian distributions of varying widths (as stated in the relevant figures) around the indicated means.

\begin{table*}
\begin{tabular}{| l | l | l | l | l | l | l | l | l | }
    \hline
    Variable & Symbol & Fig.~3a \&~3b & Fig.~3c & Fig.~3d & Fig.~4a & Fig.~4b & Fig.~5  \\ \hline
    System size & $N$ & 5 & Varied & Varied & Varied & Varied & 4  \\
    2LS splitting & $\omega_A$ & 1.8 eV & 1.8 eV & 1.8 eV & 1.8 eV & 1.8 eV & 1.8 eV \\
    2LS lifetime & $\tau_L$ & 2.5 ns & 2.5 ns & 2.5 ns & 2.5 ns & 2.5 ns & Varied  \\
    2LS separation & $r_{nn}$ & 1 nm & 1 nm & 1 nm & 1 nm & 1 nm & Varied  \\
    Equatorial angle & $\theta_{\text{eq}}$ & $\pi/2$ & $\pi/2$ & $\pi/2$ & $\pi/2$ & $\pi/2$ & $\pi/2$  \\
    Zenith angle & $\theta_{\text{zen}}$ & $\pi/4$ & $\pi/4$ & $\pi/4$ & $\pi/4$ & $\pi/4$ & Optimised   \\
    Optical suppression & $s$ & Varied & 99\% & Varied & n/a & 99\% & 99\% \\
    Extraction rate & $\gamma_x$ & $10^{-2}$ eV & $10^{-2}$ eV & $10^{-2}$ eV & n/a & $10^{-2}$ eV & $10^{-2}$ eV  \\
    Reinitialisation rate & $\gamma_r$ & Varied & Varied & n/a & n/a & $10^{-2}$ eV & $10^{-2}$ eV  \\
    Optical temperature & $T_\text{opt}$ & 5800 K & 5800 K & 5800 K & 5800 K & 5800 K & 5800 K \\
    Vibrational temperature & $T_\text{vib}$ & 300 K & 300 K & 300 K & 300 K & 300 K & 300,77,4 K  \\
    \hline
\end{tabular}
\label{tab:MTparam}
\caption{Full set of model parameters for main text figures. Additionally, photon and phonon spectral densities as discussed in Sec.~\ref{sec:diss} were employed. }
\end{table*}

\begin{table*}
\begin{tabular}{| l | l | l | l | l | l | }
    \hline
    Variable & Symbol & Fig.~S3b & Fig.~S4 GS & Fig.~S4 $||$ & Fig.~S12, S13 \& S14 \\ \hline
    System size & $N$ & 4 & 4,5 & 4,5 & 5 \\
    2LS splitting & $\omega_A$ & 1.8 eV &1.8 eV* &1.8 eV* & 1.8 eV* \\
    2LS lifetime & $\tau_L$ & 2.5 ns & 2.5 ns* & 2.5 ns* & 2.5 ns* \\
    2LS separation & $r_{nn}$ & 1 nm & 1 nm* & 1 nm* & 1 nm* \\
    Equatorial angle & $\theta_{\text{eq}}$ & Varied & $\pi/2$* & $\pi/2$* & $\pi/2$* \\
    Zenith angle & $\theta_{\text{zen}}$ & Varied & $\pi/4$* & $\pi/2$* & $\pi/4$*  \\
    Optical suppression & $s$ & 99\% & 99\% & 99.9\% & n/a \\
    Extraction rate & $\gamma_x$ & $10^{-2}$ eV & $10^{-2}$ eV & $10^{-2}$ eV & n/a \\
    Reinitialisation rate & $\gamma_r$ & $10^{-2}$ eV & $10^{-2}$ eV & $10^{-2}$ eV & n/a \\
    Optical temperature & $T_\text{opt}$ & 5800 K & 5800 K & 5800 K & n/a \\
    Vibrational temperature & $T_\text{vib}$ & 300 K & 300 K & 300 K & n/a \\
    \hline
\end{tabular}
\caption{Parameters for figures in this SI document. Figs.~\ref{fig:SpecDensPanel}-\ref{fig:polaronfig} used the same parameters as Fig.~4b of the main text, other than the deviations mentioned in their respective captions. Asterisks denote values around which Gaussian distributions of varying width were taken in disorder trials.}
\label{tab:SIparam}
\end{table*}

\begin{figure}
\flushleft{\large{\bf ~ a)}} \\
\centering
\includegraphics[width=0.4\textwidth]{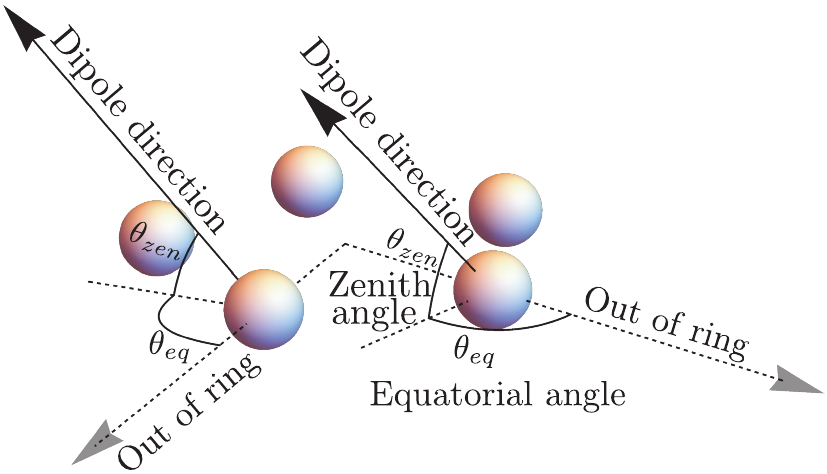} \newline
\flushleft{\large{\bf ~ b)}} \\
\includegraphics[width=0.4\textwidth]{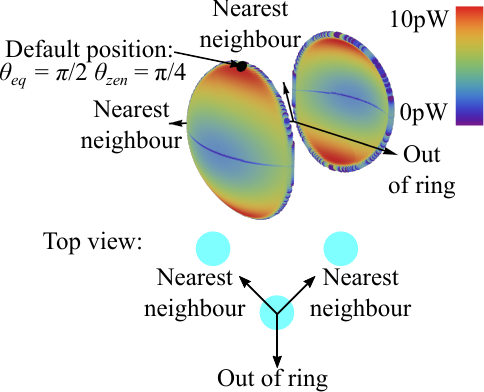}
\caption{(a) Representation in a pentamer of the equatorial angle, $\theta_{\text{eq}}$ and the zenith angle $\theta_{\text{zen}}$, used to define dipole directions. (b) Representation of how net power production of a quadmer varies as different angles are chosen for the 2LSs. Only angles which produced a positive net power are plotted. A black spot is used to denote the default angular setup used in other runs. Also included is a plan view schematic to clarify the plotted axes arrows.}
\label{fig:AngleVisual}
\end{figure}

In Fig.~\ref{fig:AngleVisual}a we include a visual representation of the two angles used to define the dipole orientations. The equatorial angle, $\theta_{\text{eq}}$, gives rotation in the plane of the ring, and the zenith angle, $\theta_{\text{zen}}$ gives inclination. The angles define a local reference for each dipole and chosen directions are constant rotations with respect to the `out-of-ring' direction, as shown.

\section{Default and optimisation of angles for Fig.~5}

As can be seen in the tables in Sec.~\ref{sec:parameters} the angles in almost all data runs are selected to be $\theta_{\text{eq}}=\frac{\pi}{2}$ and $\theta_{\text{zen}}=\frac{\pi}{4}$. Due to the nature of the large parameter scan being performed in Fig.~5 of the main text we allowed the angles to be optimised (we note that $\theta_{\text{eq}}=\frac{\pi}{2}$ is always optimal). In Fig.~\ref{fig:AngleVisual}b we demonstrate how the power output varies over a full angular scan for a quadmer run with other 2LS parameters set to default. We see the default position is close to the ideal angle, and also that the high performing region is large, a fact which proves beneficial when considering (angular) disorder in Sec.~\ref{sec:disorder}.

\section{Breakdown of parallel superabsorption ($||$-SA)}
\label{sec:parsa}

In the main text we introduce two different ring setups, $||$-SA and GS-SA. $||$-SA implements the geometry proposed in Ref.~\cite{Higgins2014aSUP}. Having all dipoles parallel maximises the total collective dipole of the system, so this would seem like the obvious choice for boosting the absorption of the system. However, as briefly mentioned in the main text, the inclusion of rapid phonon relaxation spoils the effect entirely. In the following, we show this breakdown explicitly.

\begin{figure}
\centering
	\includegraphics[width=0.45\textwidth]{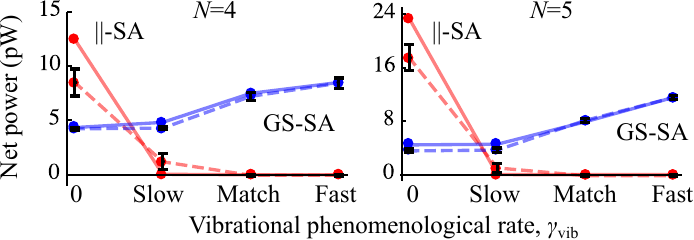}
\caption{Plots showing the net power produced by a quadmer and pentamer in $||$-SA and GS-SA setups for different phonon rates. Fixed parameters are plotted with the bold lines, while the dashed lines use 1\% input disorder over 100 trials (standard deviation error bars included). $||$-SA produces larger powers only for negligably small phonon coupling. By contrast, the GS-SA setup always produces positive net power, and only increases in performance when phonons dominate over photon processes.}
\label{fig:ParrBrkDwn}
\end{figure}

Fig.~\ref{fig:ParrBrkDwn} shows the net power produced by a quadmer and a pentamer for four different phonon rates: 
\begin{enumerate}
\item $\kappa_{\text{vib}} = 0$, i.e. no phonons; 
\item $\kappa_{\text{vib}} = 10^{-3} \times \gamma_{\text{opt}}(\omega_A) / \overline{ \omega_{\text{vib}}}$ -- `slow'; 
\item $\kappa_{\text{vib}} = \gamma_{\text{opt}}(\omega_A) / \overline{ \omega_{\text{vib}}}$ --  `match'; 
\item  $\kappa_{\text{vib}} = 10^3 \times \gamma_{\text{opt}}(\omega_A) / \overline{ \omega_{\text{vib}}}$ --  `fast'.
\end{enumerate}
Results for both $||$-SA and GS-SA are shown with red and blue lines, respectively. Solid lines use fixed input parameters, while the dashed lines are averages from 100 trials with 1\% disorder introduced over all input parameters ({\it cf.}~Sec.~\ref{sec:disorder}).

The $||$-SA case produces large net power when there is either no phonon environment at all or if phonon rates are low compared to those of optical processes. In the presence of phonons, regardless of the phonon coupling strength, $||$-SA requires some level of disorder, otherwise population gets trapped in off-ladder states from where it cannot decay. By not arriving back on the ladder it also will not be `picked up' again by the ladder-climbing reinitialisation process. The steady-state of idealised $||$-SA in the presence of any level of phonon coupling is therefore entirely on off-ladder states and away from enhanced optical transitions. For this reason the solid red line drops to zero as soon as phonons are included, whereas the dashed line (including a mild level of disorder) dips significantly but only hits zero once vibrational processes match optical ones. Both red lines tend towards zero for faster vibrational rates, demonstrating the breakdown of $||$-SA. Overall, the above discussion suggests that $||$-SA may be achievable for systems where no naturally prevalent phonon environment exists and / or where exquisite (reinitialisation) control is available, e.g.~on the superconducting circuit QED platform~\cite{Mlynek:2014aaSUP, Potocnik2018SUP}.

Meanwhile for GS-SA the power produced increases in the presence of a dominant phonon environment, both in idealised and disordered cases. GS-SA also still performs adequately even in the absence of phonons or for low phonon rates, so unlike $||$-SA it is not limited to a particular operating regime. However, as discussed in the main text, the tradeoff is that the maximum power produced from GS-SA cannot match that of $||$-SA due to its reduced collective optical dipole. GS-SA's robustness to (and even benefitting from) vibrational relaxation suggests it may be suitable for implementation across a wide range of condensed-matter nanostructures, including in the solid state and for molecular systems.

\section{Variants and extensions of the weak-coupled GS model}

\subsection{Different phonon spectral densities}
\label{sec:specdens}

In the main text, we chose an Ohmic phonon spectral density that was otherwise structureless, and in particular did not feature a high-frequency cut-off. Since the form of the phonon spectral density varies from system to system, we shall here present results when substituting this with a superohmic spectral density including a cut-off. Our results underline the robustness of the GS-SA approach and demonstrate that the results presented in the main text are generic and did not derive from the particular choice of phonon environment.

\begin{figure}
\centering
	\includegraphics[width=0.45\textwidth]{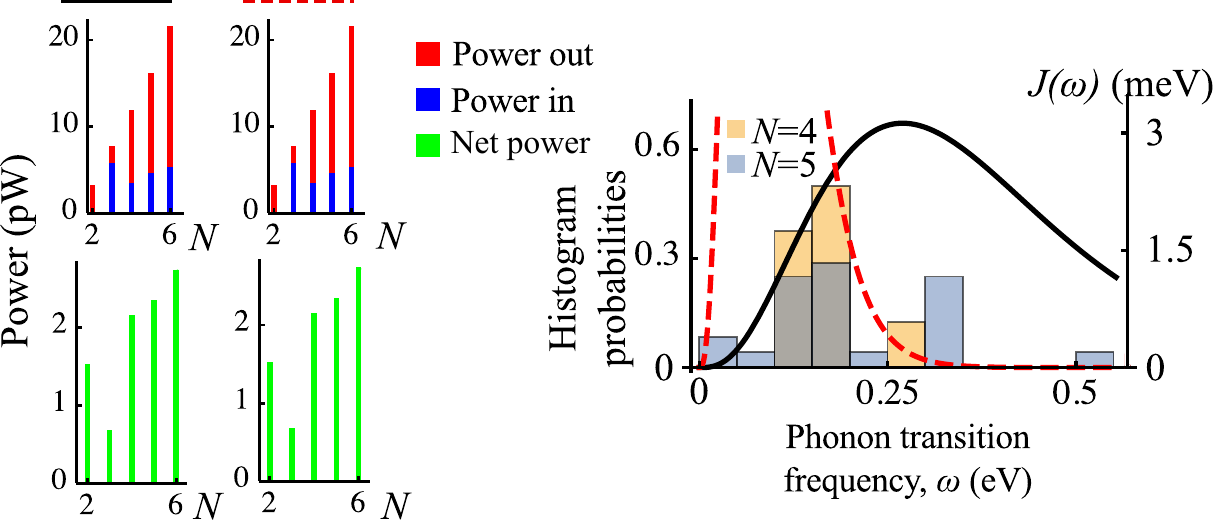}
\caption{\textbf{Left:} power in, out, and net power per site found for various systems sizes using the the spectral density of Eq.~\eqref{eqn:SpecDens}. The parameters used in the spectral density were $\lambda=5$~meV and $\omega_{\text{crit}}=90$~meV (left column, black line) and $\lambda=20$~meV and $\omega_{\text{crit}}=25$~meV (right column, red dashed line). \textbf{Right:} spectral densities overlayed on normalised histograms of the different phonon transition frequencies arising with the quadmer (yellow) and pentamer (blue). The other parameters are those of the inset of Fig.~4b in the main text.}
\label{fig:SpecDensPanel}
\end{figure}

We now consider phonon rates dependent on a spectral density in the usual way
\begin{equation}
\gamma(\omega)=J(\omega)\big(1+N(\omega)\big) ~,
\end{equation}
where $J(\omega)$ is the spectral density. Different functions could be used for the spectral density, with different forms being appropriate for different absorbing antennae. As long as the chosen phonon spectral density allows for relaxation processes to move population to the bottom of rungs, then the guide-slide concept still works. As an example we take the spectral density form
\begin{equation}
J(\omega)=\frac{\lambda\omega^3}{2\omega_{\text{crit}}^3}e^{-\frac{\omega}{\omega_{\text{crit}}}},
\label{eqn:SpecDens}
\end{equation}
\noindent where $\lambda$ denotes the reorganisation energy, and $\omega_{\text{crit}}$ is the cut-off frequency. The bar charts in Fig.~\ref{fig:SpecDensPanel} show that the guide-slide states still display superabsorbing behaviour for two different choices of $\lambda$ and $\omega_{\text{crit}}$ that are appropriate for molecular systems~\cite{Sowa2017aSUP}. The right panel plots the spectral densities and a histogram of the different phonon transition frequencies in the quadmer and pentamer systems. The guide-slide effect requires overlap of the spectral density with (at least a sufficient subset of) vibrational frequencies to allow phonon-assisted relaxation onto the ladder states. Clearly, this is the case here and results in net power output scaling super-linearly with ring size as shown in left panel of Fig.~\ref{fig:SpecDensPanel}.

\subsection{Lossy ladder-climbing reinitialisation}
\label{sec:lossylad}

In this section we employ the `lossy-ladder climbing'  reinitialisation approach. Here,  the system is over-excited in each of our reinitialisation steps, by pumping population to the top of the rung above, then allowing it to relax to the enhanced ladder state at the bottom of the rung. This `daisy-chaining' of optical excitation is followed by rapid vibrational relaxation breaks  detailed balance and validates our use of `one-way' excitation. This precludes potential concerns about reinitialisation offering a potential new loss process due to time-symmetry. Obviously, the over-excitation increased the energetic cost of reinitialising.

\begin{figure}
\centering
	\includegraphics[width=0.47\textwidth]{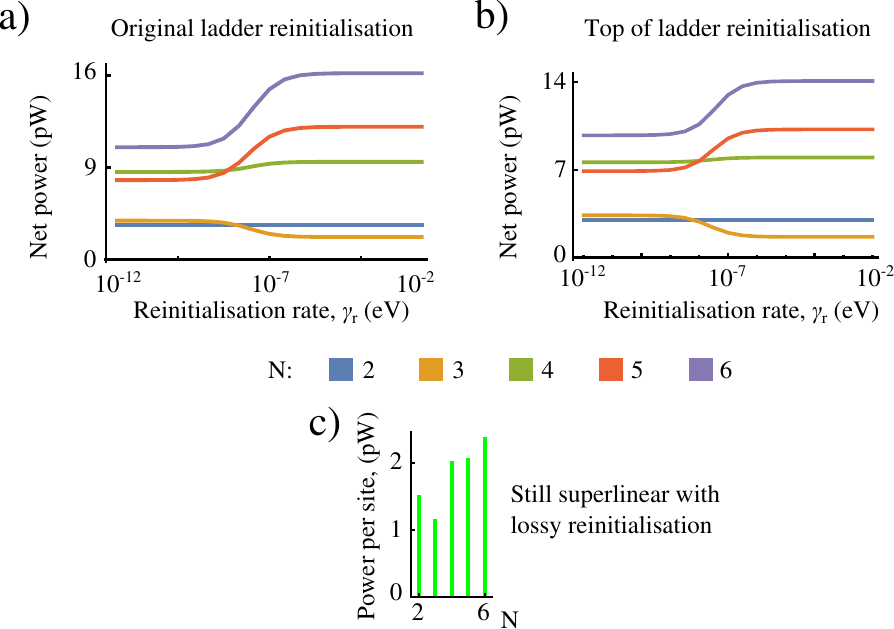}
\caption{Comparison plot of the ladder-climbinging (a) and lossy ladder-climbing (b) reinitialisation methods, both at 99\% suppression. Different system sizes are used, as well as varied reinitialisation rates. Power is slightly reduced, but the superlinear scaling of net power generation is unaffected by the change (c). Other parameters match those used in Fig.~3c of the main text.}
\label{fig:TopReinitComp}
\end{figure}

In Fig.~\ref{fig:TopReinitComp} we show a comparison of this lossy reinitalisation method against the ladder-climbing one used in Figs.~3 and 4 of the main text. We can see that the additional reinitialisation losses from using the lossy ladder climbing method do rescale the net power output, but importantly the qualitative behaviour is not affected and the superlinear scaling is maintained.

\subsection{Site-based reinitialisation}
\label{sec:siteinit}

As another variation to the more idealised ladder-climbing reinitialisation, we also consider constant pumping across all of the absorbing sites on the ring. In this case, a rapid reinitialisation rate can lead to the system occupying the top half of the ladder, causing a waste of (input) power. In contrast to ladder-climbing reinitialisation, where we use a constant and sufficiently fast reinitialisation rate for all system sizes, we trial different pumping rates, settling for the choice providing optimal net power performance. 

\begin{figure}
\centering
	\includegraphics[width=0.35\textwidth]{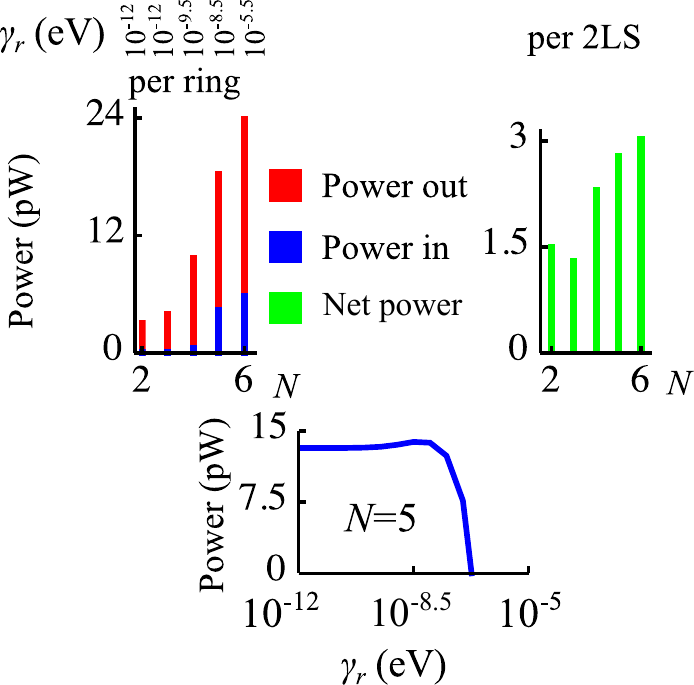}
\caption{Power input, output, and net power per site for site-based reinitialisation. In each case the reinitialisation rate which provides the optimal net power is used, the reinitialisation rate, $\gamma_r$, is shown across the top. Superlinear behaviour is unaffected by this change. Other parameters are those of the inset of Fig.~4b of the main text, except for an increased optical suppression of 99.9\%. The lower panel shows how increasing the reinitialisation rate leads to an improvement in the power out for the pentamer before pushing the system in to the top half of the ladder and wasting energy.}
\label{fig:sRiE}
\end{figure}

Fig.~\ref{fig:sRiE} qualitatively reproduces the features of Fig.~4b in the main text, and in particular retains the super-linear scaling of net power output. The inset in Fig.~\ref{fig:sRiE} shows the drop off in net power when the system reinitialisation rate becomes too fast (leading to population inversion of the individual sites and substantial steady-state population above the ladder mid-point).

\subsection{Coherent extraction process}
\label{sec:cohext}

As previously mentioned, coherent coupling between the ring and the trap increases the complexity of the calculations. However, once the right parameter regime has been identified, this does not drastically change the results. 

To implement coherent extraction, the extraction dissipator, $\mathcal{D}_x$, is removed from Eq.~\eqref{eq:fullme}, and instead terms for moving population between the ring sites and the trap are added to the Hamiltonian: $\sum_i^N\mathcal{C}_x(\hat{\sigma}_i^+\hat{\sigma}_t^-+\hat{\sigma}_i^-\hat{\sigma}_t^+)$, where $\mathcal{C}_x$ is the coupling strength between the ring and the trap. The trap site is kept degenerate with the target transition of the unperturbed ring eigenbasis (i.e. the in the absence of the trap), $\omega_t = \omega_{\text{good}}$. The new system Hamiltonian for the ring and trap is therefore

\begin{multline}
\label{eq:Hsysring}
\hat{H}_{\text{r\&t}}=\omega_A\sum_{i=1}^N \hat{\sigma}^z_i+\sum_{i,j=1}^N J_{i,j}(\vec{r}_{i,j})(\hat{\sigma}^+_i\hat{\sigma}^-_j+\hat{\sigma}^-_i\hat{\sigma}^+_j) \\
+\omega_t\hat{\sigma}^z_t+\sum_i^N\mathcal{C}_x(\hat{\sigma}_i^+\hat{\sigma}_t^-+\hat{\sigma}_i^-\hat{\sigma}_t^+)~.
\end{multline}

Resonant coherent coupling between the target ring transition and the trap implies that we no longer have unidirectional extraction and population can return from the trap back to the ring. However, for adequately fast trap decay rate $\gamma_t$, this is suppressed and does not significantly affect the results.

For relatively small $\mathcal{C}_x$ compared to $J_{i,i+1}$, the presence of the trap in the Hamiltonian can be understood as a minor perturbation of the ring eigenstates, and the trap will only resonantly extract energy from the target transition (as adjacent ladder transitions are sufficiently far detuned). However, as the coupling strength is increased, the trap and ring energy levels become hybridised. In this case the `trap' begins to also extract from other `ladder transitions' . An important ramification of losing full extraction selectivity is that the trap starts pulling the the system down the ladder, undermining the photonic bandgap suppression and counteracting the reinitialisation process. 

\begin{figure}
\centering
	\includegraphics[width=0.35\textwidth]{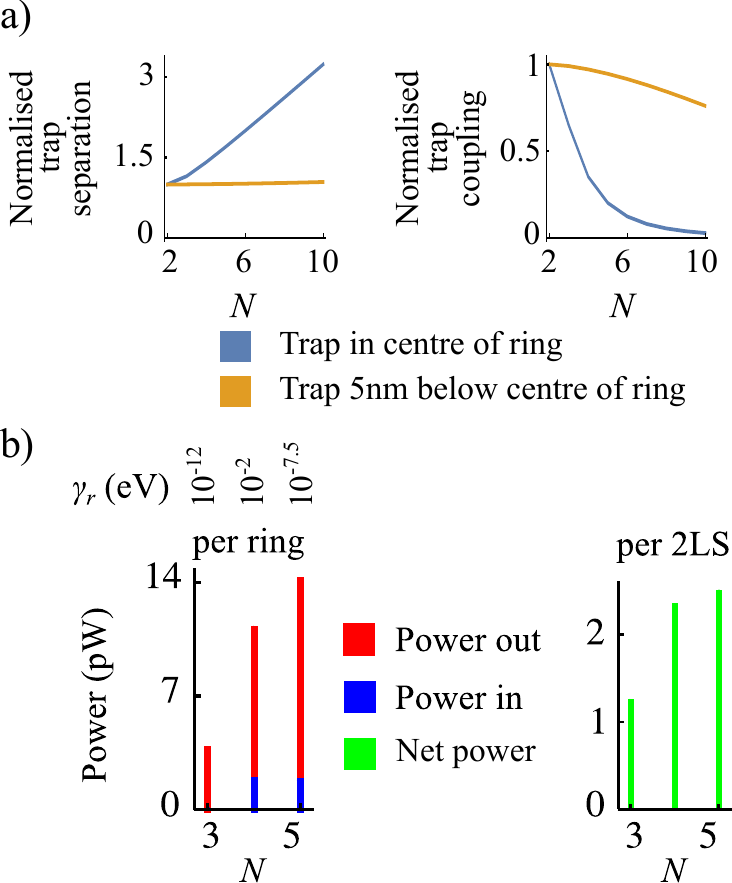}
\caption{ (a) The $N$ dependent scaling of ring-to-trap distance, and ring-to-trap coupling. (b) Power input, output, and net power per site for coherent extraction coupling to a trap in the centre of the ring. A slow base extraction rate of $10^{-4}$eV is used to keep the extraction frequency selective (see text for discussion), but is fast enough to overcome the drop off from the distant-dependent scaling. Super-linear behaviour similar to that seen in Fig.~4b of the main text is observed in this case. For this plot different choices of $\gamma_r$ were trialled, with selected rates as shown in the figure. The other parameters in these data runs are those used in Fig.~4b of the main text, except for the optical suppression which is here increased to 99.9\%.}
\label{fig:iRcE}
\end{figure}

If the coherent coupling was mediated by a dipole interaction, then one would anticipate a reduction in the coupling strength as $N$ is increased assuming the nearest-neighbour separation along the ring is kept constant; Fig.~\ref{fig:iRcE}a demonstrates how separation and coupling scale with $N$. The coupling strength drop-off is seemingly dramatic if the trap is located in the centre of the ring, but it could be reduced by positioning the trap out of the plane of the ring. For these calculations, we define our extraction coupling, $\mathcal{C}_x$, as a base rate multiplied by the appropriate $N$ dependent scaling factor.

In Fig.~\ref{fig:iRcE}b we find that superabsorbing behaviour remains possible under coherent extraction with a trap positioned at the centre of the ring (with stronger distant-dependent scaling). Here, the base extraction coupling strength of $10^{-4}$~eV is fast enough to extract most excitons absorbed by the system (for an optical bath at the solar surface temperature), while not being so strong as to break the selectivity of extraction from the target transition~\footnote{We note that calculations which did not include the distant-dependent scaling on the extraction coupling provided very similar results.}. This graph stops with the pentamer due to the substantial computational demand of solving the master equation for the case of a hexamer. The power values are not quite as high as those observed when incoherent coupling is used, as is to be expected due to the extraction speed restriction which we applied. 

Note that this plot also does not include the case of a dimer: here, letting both dipoles pointing tangentially and inclined 45$^\circ$ out of the plane of the ring leads to vanishing dipole-dipole coupling between the absorbers as the dipoles are then strictly orthogonal. This entails that for even arbitrarily weak extraction coupling the degenerate trap and dimer states are maximally hybridised and the picture of viewing the trap as a perturbation to the ring antenna fails.

\subsection{Coherent extraction \& site-based reinitialisation}
\label{sec:cohexsiteinit}

As a final variant of the model, we combine coherent extraction with site-based reinitialisation. In this setup, one needs to simultaneously ensure that the extraction coupling is not so strong that the trap starts extracting energy from other than the target transition, and that the reinitialisation proceeds fast enough to keep the system operating in the centre of the ladder, but not so fast as to push it further up.

\begin{figure}
\centering
	\includegraphics[width=0.35\textwidth]{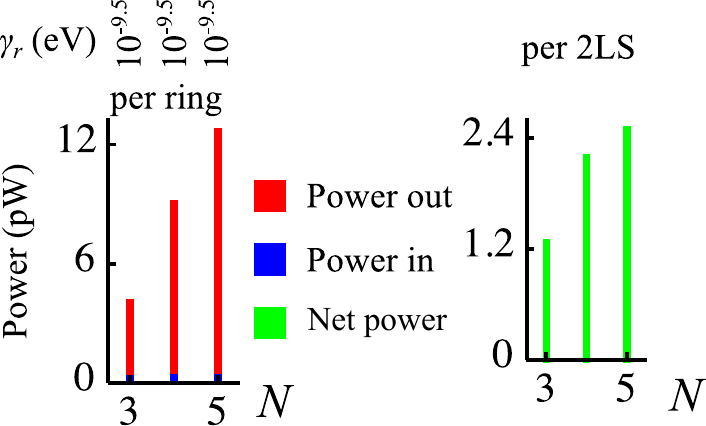}
\caption{Power input, output, and net power per site for coherent extraction and site-based reinitialisation. In each case the reinitialisation rate $\gamma_r$ was optimised to maximise net generated power. Notably, the small optimal values of $\gamma_r$ mean that the input power is also rather small. The other parameters in these data runs are consistent with those used in Fig.~4b of the main text, but again with optical suppression increased to 99.9\%.}
\label{fig:sRcE}
\end{figure}

The results of these data runs are shown in Fig.~\ref{fig:sRcE}, and once more they display superabsorbing behaviour. (Note that here we did not include an $N$-dependent scaling to the base extraction rate for these calculations, but our work in Sec.~\ref{sec:cohext} showed that for this base rate and system size there would be negligible effect on the performance by including the scaling.)

\section{Effect of non-radiative losses}

To evaluate the robustness of GS-SA to non-radiative decay processes -- of particular relevance to molecular systems -- we introduce additional dissipator terms: for each of the $N$ sites (indexed by $i$) we take an interaction matrix, $M_{\alpha,\text{nr}}=\hat{\sigma}_i^-$ and form a term via Eq.~\eqref{eq:BRdiss} with $\sigma = -1$ acting with the rate $\gamma_{nr}$. The sum over all these terms constitutes the total dissipater $\mathcal{D}_{\text{nr}}$ for non-radiative loss, assumed to act independently and equally on each site.

\begin{figure}
\centering
	\includegraphics[width=0.47\textwidth]{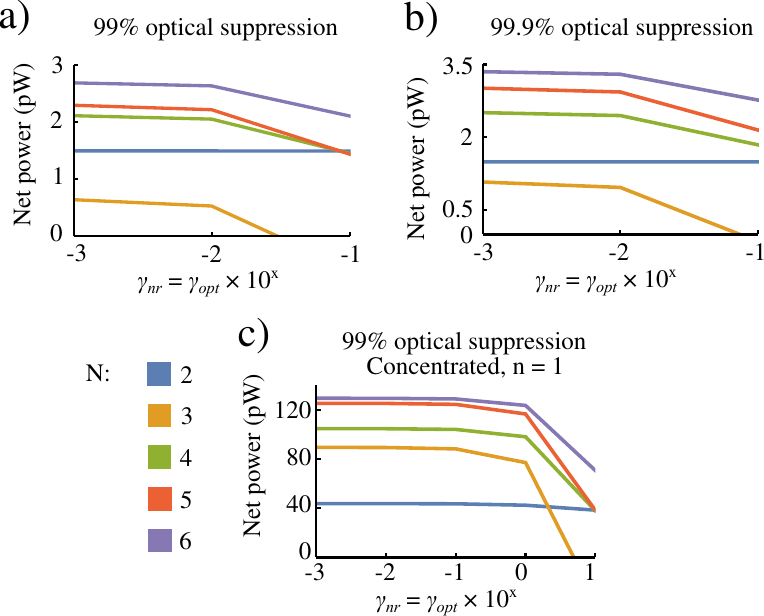}
\caption{Comparison of how the increasing the non-radiative rate, $\gamma_{nr}$ affects the net power performance of GS-SA with different system sizes. (a) and (b) All parameters match those used in Fig.~3c of the main text, with ladder-climbing reinitialisation and two strengths of optical suppression being used. (c) the optical mode occupancy is set to 1, simulating concentrated sunlight and mitigating effectively against faster $\gamma_{nr}$. } 
\label{fig:nonRadRes}
\end{figure}

In Fig.~\ref{fig:nonRadRes} we plot the optimised net power performance of GS-SA with $\mathcal{D}_{\text{nr}}$ added for the calculation of the dynamics. The rate of the non-radiative dissipator, $\gamma_{nr}$, is expressed relative to the (bare) optical decay rate, $\gamma_\text{opt}$. In Fig.~\ref{fig:nonRadRes}a  we first note that non-radiative loss has practically no effect on the dimer. For $N>3$, we find that GS-SA performance is only marginally affected when the non-radiative loss rate is substantially smaller than the radiative one. However, once it exceeds roughly 10\% we observe a rapid reduction in power output which eventually erodes any net power output before parity is reached at 99\% suppression. As shown in Fig.~\ref{fig:nonRadRes}b this decline is less dramatic if stronger suppression is available. We conclude that GS-SA will perform best for systems in which non-radiative decay is not dominant. Whilst this may be a more challenging ask on the molecular platform, examples of bright organic dyes with small~\cite{Moerner2004SUP} or even negligible~\cite{Hwang2011SUP} non-radiative decay rates exist, whilst for others such as merocyanine the ratio between radiative and non-radiative decay can to an extend be controlled through the choice of solvent~\cite{DebabrataMandal1999SUP} and substrate~\cite{Kim2012SUP}.

Further, it is worth noting the effect of dominant non-radiative decay on the GS-SA performance can be mitigated in several others ways: any modification speeding up the effective `cycle time' of the GS-SA process at unaltered non-radiative rate will prove effective. This could be due to enhancing the light matter coupling with the help of a cavity in the Purcell regime or simply by choosing brighter optical dipoles. A major and relatively straightforward improvement can be obtained from letting the system operate in a concentrated sunlight environment. Several prominent related publications have adopted this approach and chosen an optical mode occupancy $n=60000$~\cite{Creatore2013aSUP,Wertnik2018SUP}. For our purposes this is unnecessarily strong concentration, and we see enough of a performance improvement at $n=1$ to sustain GS-SA with $\gamma_{nr}$ exceeding $\gamma_{opt}$ as is typical for many organic dyes, see Fig.~\ref{fig:nonRadRes}c.

Finally, we note that non-radiative loss, being proportional to the number of excitations, scales linearly with system size, whereas the optical absorption rate and associated power output features superlinear scaling with the GS-SA approach. This suggests that larger rings will be more robust to non-radiative decay, and our calculations do indeed suggest that the trimer is worst and the hexamer least affected (c.f. Fig.~\ref{fig:nonRadRes}).

\section{Polaron transformation to strong vibrational coupling frame}

\subsection{Polaron transformation}

As coupling to vibrational modes can exceed the range of applicability of weak coupling approaches, particularly for molecular systems, we also analyse the GS effect in the polaron frame. We generally follow the approach laid out in Ref.~\cite{Nazir_2016SUP}, extended to considering an identical, strongly coupled phonon bath for each site. The transformation from the lab frame to the polaron frame (labelled with a prime) is given by
\begin{equation}
\hat{H}'=e^{\hat{S}}\hat{H}e^{-\hat{S}}~,
\end{equation}
where
\begin{equation}
\hat{S}=\sum_{i=1}^N \hat{\sigma}_i^z \otimes \sum_q \frac{g_{i,q}}{\omega_q} (\hat{b}_{i,q}^\dagger-\hat{b}_{i,q}))~.
\end{equation}
As the flat spectral phonon density used in the main paper leads to divergent expressions in this formalism~\footnote{It is well-known that the polaron transformation generally only works well with superohmic spectral densities.}, we limit ourselves to the structured spectral densities of Sec.~\ref{sec:specdens}. Transformation to the polaron frame causes a renormalisation of the system Hamiltonian which reduces the effective dipole-dipole coupling between absorbing 2LSs. Diagonalisation of the polaron frame ring Hamiltonian now produces a ladder of eigenstates with shifted energies due to renormalisation terms. 

 The dynamics of the system in the polaron frame is governed by
\begin{equation}
\frac{d}{dt}\rho'_S= -i [\hat{H}'_S,\rho'_S]+\mathcal{D}'_{\text{opt}}+\mathcal{D}'_{\text{coup}}+\mathcal{D}'_{x}+\mathcal{D}'_{r}+\mathcal{D}_{t}~,
\label{eq:polaronME}
\end{equation} 
where the effects of the vibrational bath have been absorbed into the polaron frame system Hamiltonian $\hat{H}'_S$, which includes renormalised dipole-dipole coupling terms. The transformation has removed the original exciton-phonon-interaction Hamiltonian, at the cost of introducing additional phonon-dependent dipole-dipole interaction terms with the phonon bath of the form
\begin{multline}
\sum_{i\ne j}^N J_{i,j}(\vec{r}_{i,j})\big((\hat{B}_{+,i}\hat{B}_{-,j}-B^2)\hat{\sigma}_i^+\hat{\sigma}_j^- \\
+\hat{\sigma}_i^-\hat{\sigma}_j^+(\hat{B}_{-,i}\hat{B}_{+,j}-B^2)\big).
\end{multline}
This will give rise to the phonon dissipator $\mathcal{D}'_{\text{coup}}$ above, which takes the place of $\mathcal{D}_{\text{vib}}$ from the weak coupling framework. Here, the $\hat{B}_{\pm\alpha}$ have the form ~\cite{Nazir_2016SUP} 
\begin{equation}
\hat{B}_{\pm\alpha} = \prod_q e^{\pm(\frac{g_{\alpha,q}}{\omega_q}\hat{b}_{q,\alpha}^\dagger - \frac{g^*_{\alpha,q}}{\omega_q}\hat{b}_{q,\alpha})}~,
\end{equation}
and their expectation value in the continuum limit is given by
\begin{align}
\braket{\hat{B}_{\pm,\alpha}} \equiv B = e^{-\frac{1}{2}\int_0^\infty\text{d}\omega\frac{J(\omega)}{\omega^2}\coth(\frac{\beta\omega}{2})}
\end{align}
where $\beta = (k_B T_\text{vib})^{-1}$ is the (inverse) vibrational temperature. After diagonalising the polaron transformed system Hamiltonian we proceed to create the new polaron frame vibration dissipator in the usual way~\cite{Breuer2002SUP}, obtaining: 
\begin{multline}
\mathcal{D}'_{\text{coup}}=
\sum_{n,m}w_nw_m\Big(A_m(\omega_m)\rho_S(t)A^\dagger_n(\omega_n)\Gamma_{nm}(\omega_m) \\
+ A_n(\omega_n)\rho_S(t)A^\dagger_m(\omega_m) \Gamma^\dagger_{nm}(\omega_m) \\
- \rho_S(t)A^\dagger_m(\omega_m)A_n(\omega_n)\Gamma^\dagger_{nm}(\omega_m) \\
- A^\dagger_n(\omega_n)A_m(\omega_m)\rho_S(t) \Gamma_{nm}(\omega_m)\Big) ~.
\end{multline}
As in the weak-coupling case, the weighting terms $w_{n,m}$ arise from the transformation to the diagonal basis but now they additionally include polaron frame renormalisation. The rates for each term come from the environment correlation correlation operator, $\Gamma_{nm}(\omega_m)$, of the pair of relevant environment terms in the interaction. Every possible pair will result in one of three outcomes depending on whether the paired terms pertain to the same 2LS and whether they are both of the same raising / lowering type or mixed. The four possible combinations evaluate to
\begin{align}
\braket{\hat{B}_{\pm,\alpha}(s)\hat{B}_{\pm,\alpha}(0)}_\alpha &= B^2 e^{-\phi(s)} ~,\nonumber \\
\braket{\hat{B}_{\mp,\alpha}(s)\hat{B}_{\pm,\alpha}(0)}_\alpha &= B^2 e^{+\phi(s)} ~,\nonumber \\
\braket{\hat{B}_{\pm,\alpha}(s)}_\alpha \braket{\hat{B}_{\pm,\beta}(0)}_\beta&= B^2 ~,\nonumber \\
\braket{\hat{B}_{\mp,\alpha}(s)}_\alpha \braket{\hat{B}_{\pm,\beta}(0)}_\beta&= B^2 ~,
\end{align}
where
\begin{equation}
\phi(s)=\int_0^\infty\text{d}\omega\frac{J(\omega)}{\omega^2}(\cos(\omega s)\coth(\beta\omega/2)-i\sin(\omega s)\big)~.
\end{equation}

To form the new optical dissipator, $\mathcal{D}'_\text{opt}$ we note that the form of the optical interaction becomes $\hat{B}_{+,i}\hat{\sigma}_{+}^i+\hat{B}_{-,i}\hat{\sigma}_{-}^i$, so that the vibrational dependency needs to be accounted for when creating a Bloch-Redfield dissipator. We find that
\begin{multline}
\mathcal{D}'_\text{opt}=
\sum_{n,m}\vec{d}_n\cdot\vec{d}_mP_\text{vib}\Big(A_m(\omega_m)\rho_S(t)A^\dagger_n(\omega_n)\Gamma_{nm}(\omega_m) \\
+ A_n(\omega_n)\rho_S(t)A^\dagger_m(\omega_m) \Gamma^\dagger_{nm}(\omega_m) \\
- \rho_S(t)A^\dagger_m(\omega_m)A_n(\omega_n)\Gamma^\dagger_{nm}(\omega_m) \\
- A^\dagger_n(\omega_n)A_m(\omega_m)\rho_S(t) \Gamma_{nm}(\omega_m)\Big)
\end{multline}
looks formally identical to its weak-coupling counterpart except for the the inclusion of the $P_\text{vib}$ weightings. Due to the difference in timescales between the vibrational and optical processes, the effects of the two can be separated when calculating the rates ~\cite{Nazir_2016SUP,Rouse2019SUP}. The calculation of $\Gamma_{nm}(\omega_m)$ is the same as was used in Sec.~\ref{sec:dissOPT}, though naturally it is now carried out between ladder states in the polaron frame. The value of $P_\text{vib}$ varies depending on which two terms are involved: if both pertain to the same system and are a mixed combination or raising and lowering operators then $P_\text{vib}=1$ or for equal operators $P_\text{vib}=B^4$; alternatively if they belong to different systems then regardless of operator combination $P_\text{vib}=B^2$.

The trap Hilbert space (tensored to the polaron frame ring) comes with a decay dissipator which is unaffected by the polaron frame, however, the trap energy is different as it is now tuned to the desired polaron frame ladder transition. An incoherent extraction process, like the one used in the main text, transfers population from the desired polaron frame ladder states to the excited state of the trap. The dissipator for the reinitialisation processes, $\mathcal{D}'_{r}$, uses the ladder-climbing approach described in Sec.~\ref{sec:ladderR}, but now the ladder states are in the polaron frame, and as such have a slightly different energies. An expanded discussion of the details pertaining to a multi-site polaron frame calculation can be found here~\cite{brown_2019SUP}.

As a consequence of moving to the polaron frame collective optical effects are invariably reduced, this is due to the inclusion of the vibrationally dependent $P_\text{vib}$ terms in the photon rates, and is consistent with recent results looking at a strongly coupled dimer~\cite{Rouse2019SUP}. This somewhat reduces the `brightness' of ladder states as well as the associated `darkness' of certain off-ladder states, we expect this will prove detrimental when evaluating the superabsorbing power output.

\begin{figure*}
\centering
	\includegraphics[width=0.75\textwidth]{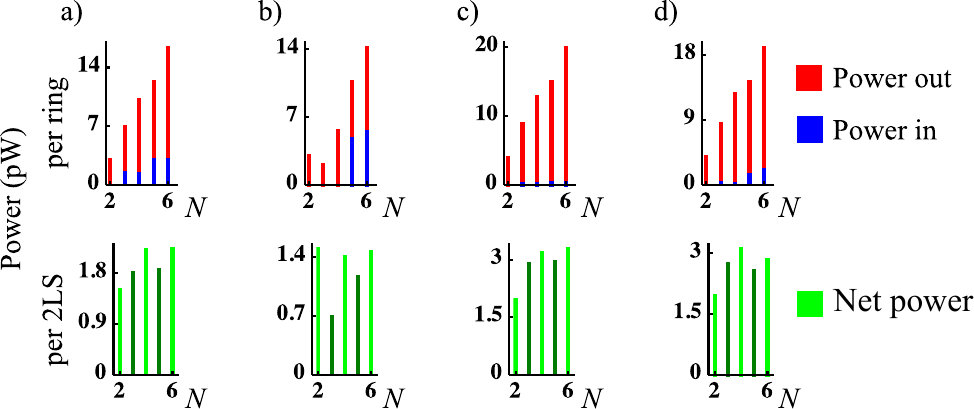}
\caption{Power input, output, and net power per site for polaron transformed GS-SA model with two different spectral densities in the conditions of the main text as well as more favourable ones summarised in Tab.~\ref{tab:POLparam}. The reinitialisation rate, $\gamma_r$, which provided optimal net power production for each system size was used. Other parameters used in the calculations are consistent with those used to produce the results in Fig.~4b of the main text.}
\label{fig:polaronfig}
\end{figure*}

Using our two phonon spectral densities we examined room temperature systems with 99\% suppression of undesired optical modes. Due to computational constraints the maximal system size we investigated was a hexamer. In Fig.~\ref{fig:polaronfig} we show the full power output data for the two cases, whilst also including results for each spectral density at cooler temperature, with stronger suppression of optical modes, and a reduced absorber lifetime of 2~ns (i.e. slightly stronger optical dipole). The parameters for all the runs are summarised in Tab.~\ref{tab:POLparam}.
\begin{table*}
\begin{tabular}{|c|c|c|c|c|c|}
\hline
Panel & Lifetime & $\lambda,\omega_{\text{crit}}$ (meV) & Suppression & Phonon temp  & Superlinear to\\
\hline
a & 2.5~ns & 5,90 & 99\% & 300~K & 6 \\
b & 2.5~ns & 20,25 & 99\% & 300~K & 2 \\
c & 2~ns & 5,90 & 99.9\% & 77~K & 6 \\
d & 2~ns & 20,25 & 99.9\% & 77~K & 4 \\
\hline
\end{tabular}
\caption{Parameters used for polaron transformed GS-SA model.}
\label{tab:POLparam}
\end{table*}

We observe that the polaron transformation noticaebly affects net power output of our antennae. Encouragingly, for weaker vibrational coupling ($\lambda=5,\omega_{\text{crit}}=20$~meV) the highest performing system was the hexamer, and there is a narrow superlinear trend when separately considering even and odd subspaces (but with the pentamer actually lying lower than the quadmer). By contrast, for the more strongly coupled vibrational environment, there was no advantage of adding more sites to a dimer based on the net power produced {\it per site}. Nonetheless, the net power of the {\it total system} did increase in both cases as sites were added. This contrasts with recently published results on the performance of collective light-harvesting systems Ref.~\cite{Hu_2018SUP} when considering extraction tuned to the lowest rung of the excitation ladder (as opposed to its centre as we do here): in that work there was an optimal $N$ beyond which the total power produced decreases. With the more favourable conditions from the lower two lines of Tab.~\ref{tab:POLparam}, the more strongly coupled $\lambda=20$~meV case performed optimally with a quadmer, while the superlinear trends in the odd and even subspaces increased in the $\lambda=5$~meV case. 

In summary, in three of the four cases considered there was at least some superlinear collective advantage to be gained from adding more sites, and in all cases the total power produced per ring monotonically increased with $N$ between trimer and hexamer as the largest studied system. Note that we did not vary the positions and orientations of the absorbers, as a full parameter scan would have been too computationally demanding, but such an investigation could have lead to a further optimisation of the results. Ideally, for intermediate vibrational coupling strengths one would adopt a  Silbey-Harris variational polaron transformation approach rather than the full-blown polaron transformation. Previous variational studies have examined a single system~\cite{McCutcheon_2011SUP}, or multiple systems, but in both cases were limited to the single excitation subspace~\cite{Pollock_2013SUP}. Whilst we expect this approach to both be superior at accurately capturing the effect of vibrations and to produce more favourable results, expanding it our multi-site and multi-excitation system is unfortunately beyond the scope of this work.

\section{Guide-slide superabsorption in presence of disorder}
\label{sec:disorder}

Finally we investigate the robustness of the GS-SA model when disorder is introduced. We look at how randomising the 2LS parameters affects the guide-slide effect by applying a normal distribution to relevant input variables: the 2LS energy splitting $\omega_A$, optical natural lifetime $\tau_L = \gamma_{\text{opt}}(\omega_A)^{-1}$, the $x, y, z$ components of the dipole positions $\vec{r}_i$, and the dipole orientation of each 2LS via $\theta_{\text{eq}}$ and $\theta_{zen}$. In an individual trial, all of these parameters are drawn from a normal distribution centred around the means given in Tab.~\ref{tab:SIparam}. 

We have performed two types of disorder calculations: (i) simulations based on the full dynamic model and calculating the net power output of a disordered GS-SA (and $||$-SA) photocell; the results are shown in Fig.~\ref{fig:ParrBrkDwn} (ii) analysing disorder in our candidate ring antennae with skewed dipoles with respect to the three criteria enabling GS-SA introduced in the main text. The former results have already been discussed, so in the following we focus on the second approach. 

\begin{figure*}
\centering
	\includegraphics[width=0.8\textwidth]{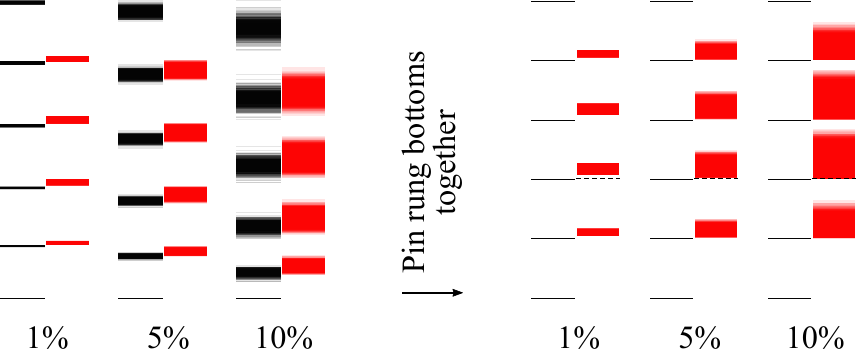}
\caption{Level structure of a pentamer with 1\%, 5\% and 10\% input disorder averaged over 1\,000 trials. The top and bottom excitation manifolds are represented by single black lines, while the other manifolds feature single black lines for their lowest energy states, and a red box covering the energy range of the remaining states in the manifold. The spacing between manifolds is not to scale with the inter-manifold spacing. On the left, one sees the increased effect of disorder in higher up manifolds (due to the zero energy rung being the same position in all of the trials). To counteract this, on the right we pin together the black lines in each manifold, making the surviving separation between these and the red boxes more apparent. The dashed black line demonstrates the energetic step which could be overcome by room temperature thermal excitation.}
\label{fig:disladder}
\end{figure*}

To address property I from the main text, Fig.~\ref{fig:disladder} shows the level structure of the skewed ring antennae under the influence of different amounts of disorder. The lowest energy in each excitation manifold, i.e. the ladder rungs, are represented by black lines, whilst all other levels in the same excitation manifold are displayed in red (for clarity offset to the right). The overall ground and fully excited states are manifolds with only a single level and thus have no associated `red boxes'. As before, the spacing between manifolds is not to scale. 

The left panel shows how disorder causes a spreading of the black lines and red boxes. Keeping the ground state energy fixed, this spread amplifies in higher excited states. A key requirement for GS-SA is that the system will always be rapidly directed to the lowest energy (ladder) state in each excitation manifold, requiring a sufficiently steep energy gradient (with steps larger than thermal energy to prevent phonon absorption). Encouragingly, there remains a clear vertical separation between black and red subset for 1\% disorder, but seemingly the lines begin to overlap for 5\% disorder, which appears to undermine property I. However, overlaying the entire ensemble is misleading in this regard: in the right panel of Fig.~\ref{fig:disladder} we pin together the ladder states for each excitation manifold~\footnote{Pinning the ladder states shows the criterion can be met individually for each member of the ensemble. Meeting it simultaneously across a disordered ensemble may be more difficult, but as shown in the next section $\omega_{\text{good}}$ and $\omega_{\text{bad}}$ remain separated so that targeted suppression and extraction remain feasible.}. This shows that the separation between black lines and red boxes survives for larger amounts of disorder. The black dashed line is elevated by 25~meV above the BTTS, so gives an indication of how likely thermal excitation away from the ladder state is at room temperature. Only a small subset of the 10\% disorder trials are prone to having the desired `guide-sliding' interfered with by thermal excitation, and in these cases a colder ambient temperature would resolve this problem. We conclude that our candidate for GS-SA continues to meet criterion I even in the presence of substantial amounts of disorder.

Fig.~4a of the main text includes disorder calculations regarding the strength of the optical transitions covering the target transition in GS-SA. This involves summing the transition strengths of the transitions from BTTS to any state in the rung above the BTTS (by the guide-slide mechanism any off-ladder population will swiftly phonon-relax back onto the ladder and be available for extraction from the TTTS). Fig.~4a considers the total useful transition strengths for different system sizes. For $N=2, \ldots, 7$, we averaged over 10\,000 trials, while the average over 1\,000, 500, and 50 trials underlies the octamer, nonamer, and decamer, respectively. For that figure of the main text, all systems parameters are drawn from a normal distribution with 5\% standard deviation.

\begin{figure}
\centering
	\includegraphics[width=0.45\textwidth]{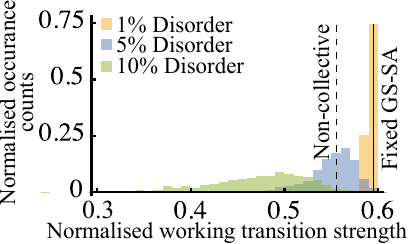}
\caption{Histograms of the maximal absorption transition strength for a pentamer with 1\%, 5\% and 10\% input disorder over 1\,000 trials. The transition strengths are normalised against the (unachievable) transition strength found in the unperturbed Dicke model. Also indicated is the no disorder GS-SA case (solid black line), and the value expected for independent absorbers which do not collectively absorb (dashed line). It is important to note from Fig.~4a in the main text that the gap between the solid line and dashed line increases with increasing $N$.}
\label{fig:distransition}
\end{figure}

In Fig.~\ref{fig:distransition} we focus on case of a pentamer and extend the 5\% disorder datapoint from Fig.~4a to 1\% and 10\% disorder. Based on 1\,000 trials, the different coloured histograms show the total transition strengths normalised against the maximum transition strength which would be expected of a pentamer in the interactionless Dicke model. We see that it takes disorder across all 2LS parameters in excess of 5\% to fully spoil the advantage of collective-enhancement. We have a further requirement to bear in mind: the most enhanced transition should indeed be the one linking the BTTS and TTTS: for 1\%, 5\% and 10\% disorder, this is the case for, respectively, 100\%, 99.7\% and 76.8\% of trials. Note that for small amounts of disorder a tiny fraction of trials exceeds the `Fixed GS-SA' line: this is since the tilting angles of the dipoles are set at a fixed value and have not been optimised, so that introducing some randomness can occasionally beat the not disordered benchmark. Fig.~\ref{fig:distransition} thus demonstrates that our candidate superabsorber also displays robustness to substantial amounts of disorder with respect to criterion II.

\begin{figure}
\centering
	\includegraphics[width=0.45\textwidth]{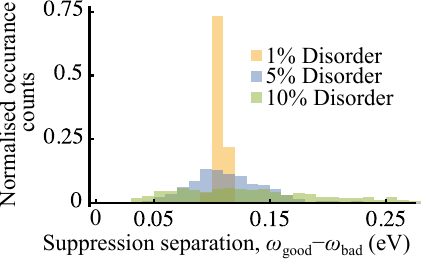}
\caption{Histograms of the frequencies used to define the optical cutoff frequency, $\omega_\text{good}$ and $\omega_\text{bad}$, for a pentamer with 1\%, 5\% and 10\% input disorder over 1\,000 trials. The optical cutoff frequency in GS-SA is defined as $\frac{\omega_\text{good}+\omega_\text{bad}}{2}$, so adequate separation is needed between these two frequencies.}
\label{fig:disGBS}
\end{figure}

Property III for GS-SA, spectral selectivity, is required to suppress coupling to certain optical modes, while allowing access to those needed to the collectively enhanced target transition. To achieve this there needs to be sufficient separation between the desired target transition frequency, $\omega_\text{good}$, and the largest undesirable frequency that needs to be suppressed, $\omega_\text{bad}$. Fig.~\ref{fig:disGBS} shows the difference between these values for differing amounts of disorder. Whilst the distribution of this gap widens with increasing amounts of disorder, reassuringly, there is no overlap and typical differences are in excess of several 10's of meV, which provides the desired spectral selectivity and meets criterion III.

In summary, this section has analysed the effect of up to 10\% disorder for a pentamer ring antenna with skewed dipoles, finding that criteria I and III continue being met for up to 10\% and criterion II is robust to at least 5\% disorder across the most relevant 2LS parameters simultaneously. Referring back to Fig.~\ref{fig:ParrBrkDwn}, we have observed that meeting the three criteria indeed translates into photocells with quantum-enhanced power generation performance. This confirms the potential of engineering and observing the GS-SA effect in real condensed-matter nanostructures.

\end{document}